\documentclass[graybox]{svmult}

\usepackage{type1cm}        % activate if the above 3 fonts are
\usepackage{makeidx}         % allows index generation
\usepackage{graphicx}        % standard LaTeX graphics tool
\usepackage{multicol}        % used for the two-column index
\usepackage[bottom]{footmisc}% places footnotes at page bottom

\usepackage{newtxtext}       % 
\usepackage{newtxmath}       % selects Times Roman as basic font
\usepackage{combelow}        % Provides the \cb{s} and \cb{t} commands for Romanian characters
\usepackage{url}
\usepackage{bm}
\usepackage{braket}          % Provides notation for bra and ket formalism
\usepackage{slashed}         % Provides Feynman slash

\definecolor{victor}{rgb}{0,0.5,0.75}
\definecolor{EW}{rgb}{0.4,0,1}

\makeindex             % used for the subject index
\begin{document}

\title*{Exact solutions in quantum field theory under rotation}
\titlerunning{Exact solutions in quantum field theory under rotation}
\author{Victor E.~Ambru\cb{s} \and Elizabeth Winstanley}
\authorrunning{Victor E.~Ambru\cb{s} \and Elizabeth Winstanley}
\institute{Victor E.~Ambru\cb{s} \at Department of Physics, West University of Timi\cb{s}oara, 
Bd.~Vasile P\^arvan 4, Timi\cb{s}oara 300223, Romania. \\ \email{Victor.Ambrus@e-uvt.ro}
\and Elizabeth Winstanley \at Consortium for Fundamental Physics, 
School of Mathematics and Statistics, University of Sheffield, 
Hicks Building, Hounsfield Road, Sheffield S3 7RH, United Kingdom. \\ \email{E.Winstanley@sheffield.ac.uk}}

\maketitle

\abstract{We discuss the construction and properties of rigidly-rotating states for free scalar and fermion fields in quantum field theory. On unbounded Minkowski space-time, we explain why such states do not exist for scalars.
For the Dirac field, we are able to construct rotating vacuum and thermal states, for which expectation values can be computed exactly in the massless case.
We compare these quantum expectation values with the corresponding quantities derived in relativistic kinetic theory.}

\section{Introduction}
\label{sec:intro}

Rigidly-rotating systems are useful toy models for studying the underlying physics of more complex rotating systems in either flat or curved space-times.
Consider a rigidly-rotating system of classical particles in flat space-time, rotating about a common axis, which we take to be the $z$-axis in the usual Cartesian coordinates.
Assuming that the particles undergo circular motion with constant angular speed $\Omega $ about the rotation axis,  
the linear speed of each particle is then $\rho \Omega $, where $\rho $ is the distance of the particle from the axis of rotation.
The speed of the particle therefore increases as the distance from the axis increases, and will become relativistic sufficiently far from the axis.
Furthermore, if $\rho $ is sufficiently large, the particle will have a speed greater than the speed of light.
Therefore a simple rigidly-rotating system cannot be realized in nature (at least in flat space-time), and the system must either be bounded in some way to prevent superluminal speeds, or else the system cannot be rigidly-rotating.

Although unbounded rigidly-rotating systems cannot be realized in flat space-time, nonetheless the study of rigidly-rotating systems in both relativistic kinetic theory (RKT) and quantum field theory (QFT) has a long history.
The simplicity of the system allows many quantities of physical interest (such as quantum expectation values) to be computed exactly, which enables the extraction of the underlying physics. 
Many deep physical properties of rotating systems have been revealed by this approach, and in this chapter we outline some of the most important.

Our motivation for studying rigidly-rotating systems in QFT comes from both astrophysics and heavy ion collisions.
In astrophysics, rigid rotation can be
induced near rapidly-rotating magnetars or in accretion disks 
around black holes, where the field close to the surface of the star is sufficiently strong 
to lock charged particles into magnetically dominated accretion flow.
The superluminal motion of the 
plasma constituents can be prohibited by the bending of the magnetic 
field lines far from the axis of rotation \cite{Meier:2012}.
Particle 
geodesics on rotating black hole space-times also exhibit rigid rotation close to the event horizon
due to the frame-dragging effect \cite{Chandrasekhar:1985kt}. 
Quantum effects are important for black holes, which emit thermal quantum radiation \cite{Hawking:1974sw}.
Whether or not it is possible to define a quantum state representing a quantum field in thermal equilibrium with a rotating black hole depends on whether one considers a scalar field (in which case such a state does not exist \cite{Frolov:1989,Kay:1988mu}) or a fermion field (where a state can be constructed, but is divergent far from the black hole \cite{Casals:2013}). 

In the context of strongly interacting systems, rigid rotation can occur 
in the quark-gluon plasma (QGP) formed in the early stages following 
the collision of (ultra-)relativistic heavy ions \cite{Jacak:2012}. 
Just as 
a magnetic field can induce a charge current along the magnetic field direction 
in fermionic matter through the chiral magnetic effect,  rigid rotation 
can induce an axial current through an analogous chiral vortical effect 
(CVE)
\cite{Kharzeev:2016}. 
Due to the latter,
the rotating fluid 
becomes polarised along the rotation axis. 
This polarisation was recently 
demonstrated through measurements of the properties of the 
decay products of $\Lambda$-hyperons \cite{STAR:2017,STAR:2018}.
Interest in studying the properties of rigidly-rotating quantum systems 
has surged in the past few years, with recent studies addressing 
the hydrodynamic description of fluids with spin \cite{Florkowski:2018},
the role of the spin tensor in nonequilibrium thermodynamics \cite{Becattini:2019}
and the properties of thermodynamic equilibrium for the free Dirac field 
with axial chemical potential \cite{Buzzegoli:2018}.

Our focus on this chapter is rigidly-rotating systems in flat-space QFT. 
We consider the simplest types of quantum field, namely a free scalar or Dirac fermion field.
By ignoring the self-interactions of the quantum field, and the curvature of space-time, we are able to study in detail the effect of rotation alone.
The construction of rotating vacuum and thermal states for these fields is compared with the corresponding construction of nonrotating vacuum and thermal states.
Here the difference between bosonic and fermionic quantum fields play a major role.
Having constructed the rotating states, we then elucidate their physical properties by studying, for the fermion field, the expectation values of the fermion condensate (FC), charge current (CC), axial current (AC), and stress-energy tensor (SET).
We compare these with the analogous quantities computed within the framework of RKT, to elucidate the effects of quantum corrections.

This chapter is structured as follows.
The problem of rigid rotation at finite temperature is addressed 
from an RKT perspective in section~\ref{sec:RKT}.
Section \ref{sec:therm} considers the construction of rigidly-rotating states in QFT, showing in particular that these states do not exist for a free quantum scalar field on unbounded flat space-time. 
The rest of the chapter is therefore devoted to the free Dirac field only.
Mode solutions of the Dirac equation are derived with 
respect to a cylindrical coordinate system in section \ref{sec:modes}.
We briefly consider nonrotating thermal expectation values (t.e.v.s) in section \ref{sec:tevs0}, and demonstrate that there are no quantum corrections for these states.
On the other hand, for rotating states, the t.e.v.s  constructed in section \ref{sec:tevs} are modified in QFT compared to the RKT results.
We examine the physical properties of these quantum corrections for the SET in particular in section \ref{sec:SET}.
The above discussion has focussed on unbounded flat space-time, and we briefly review some more general scenarios in section \ref{sec:further} before presenting our conclusions in section \ref{sec:conc}.

\section{Relativistic kinetic theory}
\label{sec:RKT}

Before we address the properties of rigidly-rotating systems in QFT, we first consider the RKT perspective.
We briefly describe
the main features of a distribution of Bose-Einstein or Fermi-Dirac particles 
in global thermal equilibrium (GTE) undergoing rigid rotation.

\subsection{Rigidly-rotating thermal distribution}
\label{sec:RKT:cyl}

Consider particles of mass $M$ and four-momentum $p^{\mu }$ in GTE in the absence of external forces.
The configuration of particles is described by the distribution function $f$, which satisfies the
relativistic Boltzmann equation \cite{Cercignani:2002}
\begin{equation}
 p^\mu \partial_{\mu} f = {\mathcal {C}}[f],
 \label{eq:boltz}
\end{equation}
using Cartesian coordinates on Minkowski space-time, so that 
$x^{\mu} = (t, x, y, z)^T$. 
In~\eqref{eq:boltz}, ${\mathcal {C}}[f]$ is the collision 
operator, which drives the fluid towards local thermal equilibrium and whose properties give the form of the equilibrium distribution function.
For neutral scalar particles, the equilibrium is described by the
Bose-Einstein distribution function
\begin{equation}
    f_{{\rm {S}}} = \frac{g_{\rm {S}}}{\left(2\pi \right) ^{3}} \left[
    \exp \left( p_{\lambda } \beta ^{\lambda }\right) - 1
    \right] ^{-1},
    \label{eq:fscalar}
\end{equation}
where $g_{\rm {S}}$ is the number of bosonic degrees of freedom and
$\beta ^{\mu  } = u^{\mu }/T$ is the four-temperature, 
with $T$ the local temperature and $u^{\mu }$ the four-velocity.
For simplicity, we
do not include a chemical potential in the scalar case.
The Fermi-Dirac distribution function, including a local chemical potential $\mu $ is
\begin{equation}
 f_{\rm {F}} = \frac{g_{\rm {F}}}{(2 \pi)^3} 
 \left[\exp\left( p_{\lambda } \beta ^{\lambda  }- \mu/T \right) + 1\right]^{-1},
 \label{eq:ffermion}
\end{equation}
where $g_{\rm {F}}$ is a degeneracy factor taking into account internal 
degrees of freedom, such as spin and colour charge. 

GTE
is achieved when the distribution function
(\ref{eq:fscalar}, \ref{eq:ffermion}) satisfies the Boltzmann equation \eqref{eq:boltz}. 
The fluid can be in GTE only when
\begin{equation}
 \partial_\lambda(\mu /T) = 0, \qquad 
 \partial_\lambda \beta_\kappa + \partial_\kappa \beta_\lambda = 0.
 \label{eq:globalT}
\end{equation}
The first equality implies that, in the fermion case, the chemical 
potential is proportional to the temperature. 
The second 
equation requires that the four-temperature $\beta^\mu$ is a Killing vector. 
For Minkowski space-time, the general solution of the Killing equation 
allows $\beta_\mu$ to be written in the form:
\begin{equation}
 \beta_\mu = b_\mu + \varpi_{\mu\nu} x^\nu,
 \label{eq:RKT_killing}
\end{equation}
where the four-vector $b^\mu$ and the thermal 
vorticity tensor $\varpi_{\mu\nu} = -\frac{1}{2}
(\partial_\mu \beta_\nu - \partial_\nu \beta_\mu)$ are constants 
in GTE.

In order to describe a state of rigid rotation with 
angular velocity 
$
\bm{\Omega} = \Omega \bm{k}
$
about the $z$-axis, the constants appearing in 
\eqref{eq:RKT_killing} can be taken to be:
\begin{equation}
 b^\mu = T_{0}^{-1}\delta^\mu{}_0 , \qquad
 \varpi_{\mu\nu} = \Omega T_{0}^{-1} \left(
 \eta_{\mu x} \eta_{\nu y} - \eta_{\mu y} \eta_{\nu x} \right),
\end{equation}
where $\eta _{\mu \nu } = {\rm {diag}} (1, -1, -1, -1)$ is the usual Minkowski metric.
These values correspond to the four-temperature 
$
\beta^\mu = T^{-1}_{0}(1, -\Omega y, \Omega x, 0),
$
where the physical interpretation of the constant $T_{0}$ is discussed below.
Since the rigidly-rotating state is invariant under 
rotations about the $z$-axis, it is
convenient to employ cylindrical coordinates 
$x^\mu = (t, \rho, \varphi, z)$ to refer to various 
vector or tensor components. 
Using the standard 
transformation formulae for vector components yields:
\begin{equation}
 \beta^t = T_{0}^{-1}, \qquad 
 \beta^\rho = 0, \qquad 
 \beta^{\varphi} = \Omega T_0^{-1}, \qquad 
 \beta^z = 0.
 \label{eq:beta_cyl}
\end{equation}

In our later discussion, it will prove useful to express  vector and tensor components of physical quantities
relative to an orthonormal (non-holonomic) tetrad $\{e_{\hat{\alpha}}\}$ 
consisting of four mutually orthogonal vectors of unit norm, 
$e_{\hat{\alpha}} = e_{\hat{\alpha}}^\mu \partial_\mu$, 
defined as:
\begin{equation}
 e_{\hat{t}} = \partial_t, \qquad 
 e_{\hat{\rho}} = \partial_\rho, \qquad 
 e_{\hat{\varphi}} = \rho^{-1} \partial_\varphi, \qquad 
 e_{\hat{z}} = \partial_z,
 \label{eq:frame_cyl}
\end{equation}
which satisfy the orthogonality relation:
\begin{equation}
 g_{\mu\nu} e^\mu_{\hat{\alpha}} e^\nu_{\hat{\sigma}} =
 \eta_{\hat{\alpha} \hat{\sigma}},
\end{equation}
where $g_{\mu\nu} = {\rm diag}(1, -1, -\rho^2, -1)$ is the 
metric tensor of Minkowski space-time with respect to the cylindrical coordinates.

Writing the four-temperature \eqref{eq:beta_cyl} with respect to the 
tetrad \eqref{eq:frame_cyl} yields the tetrad components:
\begin{equation}
 \beta^{\hat{\alpha}} = \eta ^{\hat {\alpha }{\hat {\sigma }}}e_{\hat{\sigma  }}^\mu \beta_\mu 
 = T_{0}^{-1}(1, 0, \rho \Omega, 0).
\end{equation}
The squared norm of the above expression can be obtained 
using either the coordinate components $\beta^\mu$ or 
the tetrad components $\beta^{\hat{\alpha}}$, as follows:
\begin{equation}
 \beta^2 = g_{\mu\nu} \beta^\mu \beta^\nu = 
 \eta_{\hat{\alpha} \hat{\sigma}} \beta^{\hat{\alpha}} 
 \beta^{\hat{\sigma}} = T_0^{-2} (1 - \rho^2 \Omega^2).
\end{equation}
Since $\beta^{\hat{\alpha}} = u^{\hat{\alpha}}/T$
and the four-velocity $u^{\hat{\alpha}}$ has unit 
norm by definition, it can be seen that the quantity 
$\sqrt{\beta^2}$ is the local inverse temperature $T^{-1}$.
For a rigidly rotating system, the four-velocity has the tetrad components:
\begin{equation}
u^{\hat{\alpha}} = \Gamma(1, 0, v^{\hat{\varphi}}, 0) ,
\label{eq:4velocity}
\end{equation}
where we find the following relations:
\begin{equation}
 T= T_{0}\Gamma, \qquad 
 v^{\hat{\varphi}} = \rho \Omega, \qquad 
 \Gamma = (1 - \rho^2 \Omega^2)^{-1/2},
 \label{eq:RKT_sol}
\end{equation}
where $T$ is the local temperature.
Equation~\eqref{eq:RKT_sol} shows that $T_{0}$ 
is the temperature on the rotation axis and away from the axis the local temperature increases linearly with the Lorentz factor $\Gamma $ characterising the rigid rotation. 
Furthermore,  we can readily identify 
the speed-of-light surface (SLS), which is the surface where the fluid rotates at the speed of light:
\begin{equation}
 \rho_{\rm SLS} = \Omega^{-1}.
\end{equation}
As expected, the Lorentz factor $\Gamma$ diverges on the SLS, and 
so does the local temperature $T$.
Starting from the velocity field in \eqref{eq:4velocity}, it is possible to compute the kinematic vorticity\footnote{We use the convention that $\varepsilon^{\hat{0}\hat{1}\hat{2}\hat{3}} = \varepsilon^{\hat{t}\hat{\rho}\hat{\varphi}\hat{z}} = 1$.}, $\omega^{\hat{\alpha}} = \frac{1}{2} \varepsilon^{\hat{\alpha}\hat{\beta}\hat{\gamma}\hat{\sigma}} u_{\hat{\beta}} \nabla_{\hat{\gamma}} u_{\hat{\sigma}}$, acceleration, $a^{\hat{\alpha}} = u^{\hat{\beta}} \nabla_{\hat{\beta}} u^{\hat{\alpha}}$ and circular vector 
$\tau^{\hat{\alpha}} = -\varepsilon^{\hat{\alpha}\hat{\beta}\hat{\gamma}\hat{\sigma}} \omega_{\hat{\beta}} a_{\hat{\gamma}} u_{\hat{\sigma}}$ \cite{Becattini:2015,Becattini:2015prd,Ambrus:2019ayb,Ambrus:2019khr}:
\begin{equation}
 \omega^{\hat{\alpha}} = \Gamma^2 \Omega (0,0,0,1), \quad 
 a^{\hat{\alpha}} = -\rho \Gamma^2 \Omega^2 (0,1,0,0), \quad 
 \tau^{\hat{\alpha}} = -\rho \Omega^3 \Gamma^5 (\rho \Omega, 0, 1, 0).
 \label{eq:kinematic}
\end{equation}

\subsection{Macroscopic quantities} 
\label{sec:RKT:macro}

At sufficiently high temperatures, pair production processes can occur. 
It is thus necessary to account for the presence of both particle and 
anti-particle species. 
We consider only the simplest model. 
For the neutral scalar field in thermal equilibrium, particles and anti-particles have the same distribution function $f_{\rm {S}}$ \eqref{eq:fscalar}.
Fermions and anti-fermions are distributed 
according to the Fermi-Dirac distribution \eqref{eq:ffermion} at the same  temperature 
$T$ and macroscopic velocity $u^{\hat{\alpha}}$, while the chemical 
potential is taken with the opposite sign for anti-particles:
\begin{equation}
 f_{q/\overline{q}} = \frac{g_{\rm {F}}}{(2 \pi)^3} 
 \left[\exp(p_\lambda \beta ^\lambda \mp \mu/T) + 1\right]^{-1},
 \label{eq:RKT_feq_rot}
\end{equation}
where $f_{q}$ is the distribution for fermions and $f_{\overline {q}}$ that for anti-fermions.
For the rigidly-rotating system, the contraction of the four-temperature $\beta^\mu$ with the particle four-momentum is:
\begin{equation}
 p_{\lambda } \beta ^{\lambda } = T_{0}^{-1}
 \left[ p^t - \bm{\Omega} \cdot (\bm{x} \times \bm{p})\right]
 = T_{0}^{-1}(p^t - \Omega M^z)
 = T_{0}^{-1} {\widetilde {p}}^{t}
\end{equation}
where $M^z$ denotes the $z$ component of the angular momentum, and we have defined the co-rotating energy ${\widetilde {p}}^{t}$ by 
\begin{equation}
 \widetilde{p}^t = p^t -\Omega M^z.
 \label{eq:prot}
\end{equation}

We first consider the zero temperature limit.
From \eqref{eq:fscalar}, it is clear that the scalar distribution function $f_{\rm {S}}\rightarrow 0 $ as $T_{0}\rightarrow 0$, as expected. 
The situation is more complicated for the fermion distribution function \eqref{eq:RKT_feq_rot}, and depends on the sign of $p_{\lambda }\beta ^{\lambda }\pm \mu /T$.
Noting that $\mu /T=\mu _{0}/T_{0}$ (where $\mu _{0}$ is the chemical potential on the axis of rotation) is a constant from \eqref{eq:globalT}, 
the zero temperature limit of \eqref{eq:RKT_feq_rot} is:
\begin{equation}
 \lim_{T_{0} \rightarrow 0} f_{q/\overline{q}} =
 \frac{g_{\rm {F}}}{(2\pi)^3} 
 \Theta(\pm {\mathcal {E}}_F - \widetilde{p}^t), \qquad 
 \label{eq:RKT_EF}
\end{equation}
where ${\mathcal {E}}_F = \mu _{0}$ is the Fermi level 
and $\Theta $ is the Heaviside step function, equal to one when its argument is positive and zero otherwise.
Thus, the particle/anti-particle distributions have 
non-vanishing values only when 
$\widetilde{p}^t < \mu_0$ for particles and $\widetilde{p}^t < -\mu_0$ for anti-particles.

Starting from the distribution functions, we can define the SET $T^{\hat{\alpha}\hat{\sigma}}_{\rm {S}/{\rm {F}}}$ for either a scalar or fermion field as follows:
\begin{equation}
T_{\rm {S}}^{\hat{\alpha}\hat{\sigma}} 
=\int \frac{d^3p}{p^{\hat{t}}} 
 p^{\hat{\alpha}} p^{\hat{\sigma}} f_{S}, \qquad
T_{\rm {F}}^{\hat{\alpha}\hat{\sigma}} = \int \frac{d^3p}{p^{\hat{t}}} 
 p^{\hat{\alpha}} p^{\hat{\sigma}}
 \left[f_q + f_{\overline{q}} \right] .
 \label{eq:SETRKT}
\end{equation}
For the fermion field, we can also define
the macroscopic CC
$J^{\hat{\alpha}}$:
\begin{equation}
 J^{\hat{\alpha}} = \int \frac{d^3p}{p^{\hat{t}}} p^{\hat{\alpha}} 
 [f_q - f_{\overline{q}}].
 \label{eq:CCRKT}
 \end{equation}
By construction, $J^{\hat{\alpha}}$ and $T^{\hat{\alpha}\hat{\sigma}}_{\rm {S}/{\rm {F}}}$ are 
space-time tensors. 
Due to the structure of the scalar and  Fermi-Dirac distributions,
the free indices of these quantities can be carried only by the Minkowski 
metric tensor $\eta^{\hat{\alpha}\hat{\beta}}$ or the macroscopic 
velocity $u^{\hat{\alpha}}$. 
These simple considerations immediately 
imply the perfect fluid form for the CC and SET:
\begin{equation}
 J^{\hat{\alpha}} = Q_{{\rm {F}}} u^{\hat{\alpha}}, \qquad
 T_{{\rm {S}}/{\rm {F}}}^{\hat{\alpha}\hat{\sigma}} = (E_{\rm {S}/{\rm {F}}} + P_{\rm {S}/{\rm {F}}}) u^{\hat{\alpha}} u^{\hat{\sigma}} - P_{\rm {S}/{\rm {F}}} \eta^{\hat{\alpha}\hat{\sigma}},
 \label{eq:RKT_SET}
\end{equation}
where $Q_{\rm {F}}$ is the fermion charge density, $E_{\rm {S}/{\rm {F}}}$ is the energy density and $P_{\rm {S}/{\rm {F}}}$ is the pressure.
An expression can be obtained for $Q_{\rm {F}}$ by contracting $J_{\rm {F}}^{\hat{\alpha}}$
with $u_{\hat{\alpha}}$. 
Similarly, $E_{\rm {S}/{\rm {F}}}$ is obtained by contracting 
$T_{\rm {S}/{\rm {F}}}^{\hat{\alpha}\hat{\sigma}}$
with $u_{\hat{\alpha}} u_{\hat{\sigma}}$, while a contraction of \eqref{eq:RKT_SET} 
with $\eta_{\hat{\alpha}\hat{\sigma}}$ yields the combination $E_{\rm {S}/{\rm {F}}} - 3P_{\rm {S}/{\rm {F}}}$ on the right hand 
side. The above procedure applied to $Q_{\rm {F}}$ yields:
\begin{equation}
 Q_{\rm {F}} = \frac{g_{\rm {F}}}{(2\pi)^3} \int \frac{d^3p}{p^{\hat{t}}} \left( u^{\lambda }  p_{\lambda } \right) 
 \left(\frac{1}{e^{\left( u^{\lambda }p_{\lambda } - \mu \right) /T } + 1} - 
 \frac{1}{e^{\left( u ^{\lambda } p_{\lambda } + \mu \right) /T} + 1}\right).
\end{equation}
Taking advantage of the Lorentz invariance of the integration 
measure $d^3p / p^{\hat{t}}$, a Lorentz transformation can 
be performed on $p^\lambda $ such that $p^{\lambda }u_{\lambda  } = p^{\hat{t}}$.
Switching to spherical coordinates in momentum space, the 
integral over the angular coordinates is straightforward and gives 
\begin{equation}
Q_{\rm {F}} = \frac{g_{\rm {F}}}{2\pi^2} \int_0^\infty dp\, p^2 \left(
 \frac{1}{e^{\left( p^{\hat{t}} - \mu \right)/T}  + 1} - 
 \frac{1}{e^{\left( p^{\hat{t}} + \mu \right)/T} + 1} \right)  ,
 \label{eq:RKT_QF}
\end{equation}
where $p=\left| {\bm {p}} \right| $ is the magnitude of the three-momentum.
Similarly, we find, for the scalar field,
\begin{equation}
    \begin{pmatrix}
    E_{\rm {S}} \\ E _{\rm {S}} - 3P_{\rm {S}}
    \end{pmatrix}
    = \frac{g_{\rm {S}}}{2\pi ^{2}} \int _{0}^{\infty } \frac{ p^{2} dp}{p^{\hat {t}}}
    \begin{pmatrix}
    ( p^{{\hat {t}}}) ^{2} \\ M^{2}
    \end{pmatrix}
    \frac{1}{e^{ p^{\hat {t}} /T } - 1} ,
    \label{eq:RKT_gen_S}
\end{equation}
while for the fermion field we have
\begin{equation}
 \begin{pmatrix}
  E_{\rm {F}} \\ E_{\rm {F}} - 3P_{\rm {F}}
 \end{pmatrix}= \frac{g_{\rm {F}}}{2\pi^2} \int_0^\infty \frac{p^2 dp}{p^{\hat{t}}}
 \begin{pmatrix}
  (p^{\hat{t}})^2 \\ M^2
 \end{pmatrix}
 \left(\frac{1}{e^{\left( p^{\hat{t}} - \mu \right) /T} + 1} + 
 \frac{1}{e^{\left( p^{\hat{t}} + \mu \right) /T} + 1} \right).
 \label{eq:RKT_gen_F}
\end{equation}
Since the integrands above exhibit exponential decay at large values of 
$p$, they are amenable to numerical integration.
The expressions (\ref{eq:RKT_QF}, \ref{eq:RKT_gen_S}, \ref{eq:RKT_gen_F}) remain valid if the system is stationary rather than rotating, in which case $T=T_{0}$ and $\mu =\mu_0$ are constants.

In the massless limit, $p^{\hat{t}} = p$, $E_{\rm S/F} = 3P_{\rm S/F}$ and the integrals in 
(\ref{eq:RKT_QF}, \ref{eq:RKT_gen_S}, \ref{eq:RKT_gen_F}) can be performed analytically 
\cite{Jaiswal:2015mxa}, giving the 
charge density $Q_F$ and pressures $P_{\rm S}$ and $P_{\rm F}$ as:
\begin{equation}
 Q_{\rm {F}} = 
 \frac{g_{\rm F} \mu}{6} \left(T^2 + \frac{\mu^2}{\pi^2}\right),\quad
    P_{\rm {S}} = \frac{\pi ^{2} g_{\rm {S}} T^4}{90},
    \quad 
    P_{\rm {F}} = \frac{7\pi^2 g_{\rm {F}} T^{4}}{360} 
    + \frac{g_{\rm F} T^2 \mu^2}{12} 
    + \frac{g_{\rm F} \mu^4}{24\pi^2},
 \label{eq:RKT_M0}
\end{equation}
where $\mu = \mu_0 \Gamma$ and $T = T_0 \Gamma$.
We also compute the massless limit of the 
ratio $T^\mu{}_\mu / M^2 = (E - 3P) / M^2$:
\begin{equation}
 \lim_{M \rightarrow 0} \frac{E_S - 3P_S}{M^2} = 
 \frac{g_{\rm {S}} T^2}{12}, \qquad 
 \lim_{M \rightarrow 0} \frac{E_F - 3P_F}{M^2} = 
  \frac{g_{\rm {F}} T^2}{12} + 
 \frac{g_{\rm F} \mu^2}{4 \pi^2}.
 \label{eq:RKT_M0tr}
\end{equation}
The Lorentz factor $\Gamma $ (\ref{eq:RKT_sol}) 
and thus $\mu$ and $T$ diverge 
as $\rho \rightarrow \Omega ^{-1}$ and the SLS is approached. 
Therefore, for massless particles, all macroscopic quantities are divergent on the SLS.
Including the chemical potential does not alter the rate at which the quantities diverge, but does increase their values on the axis of rotation. 

To understand the effect of the particle mass, the integrals in (\ref{eq:RKT_QF}, \ref{eq:RKT_gen_S}, \ref{eq:RKT_gen_F})  are performed numerically.
The resulting quantities depend on the angular speed $\Omega $, the temperature on the axis $T_{0}$, the chemical potential on the axis $\mu _{0}$, the particle mass $M$, the distance from the axis $\rho $ and the numbers of degrees of freedom (dof) $g_{\rm {S}}$, $g_{\rm {F}}$.
Here we consider values of these parameters which are pertinent for the QGP formed in heavy ion collisions.
An analysis of the QGP fluid produced in accelerators indicates that it has the greatest vorticity of any fluid produced in a laboratory 
\cite{STAR:2017,Wang:2017}, with $\hbar \Omega \simeq 6.6\ {\rm MeV}$, where $\hbar $ is the reduced Planck's constant. 
For this value of $\Omega $, the SLS is located at $c/\Omega \simeq 30\ {\rm {fm}}$, roughly twice the size of a gold nucleus.
For the temperature, we consider a typical value for heavy ion collisions of $k_{\rm {B}} T_{0} \simeq 0.2\ {\rm GeV}$ \cite{STAR:2017}, where $k_{B}$ is the Boltzmann constant.
In the relativistic collision of gold nuclei, 
a typical value of the chemical potential is
$\mu _{0} \simeq 0.1\ {\rm GeV}$ \cite{Huang:2012}.
For the particle mass $Mc^{2}$, we consider the pion mass
($0.140\ {\rm GeV}$), the $\rho$ meson mass ($0.775\ {\rm GeV}$),
the $\Lambda^0$-hyperon mass ($1.116\ {\rm GeV}$) 
and the $\Lambda^+_c$-charmed hyperon mass ($2.286\ {\rm GeV}$)
\cite{particlerev}.\footnote{Note that, since mesons are bosons, the Fermi-Dirac statistics cannot be strictly applied.}

\begin{figure}
    \centering
\begin{tabular}{cc}
 \includegraphics[width=0.45\linewidth]{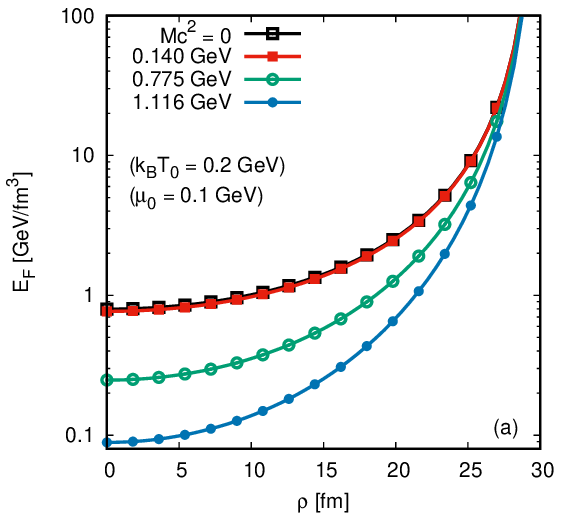} &
 \includegraphics[width=0.45\linewidth]{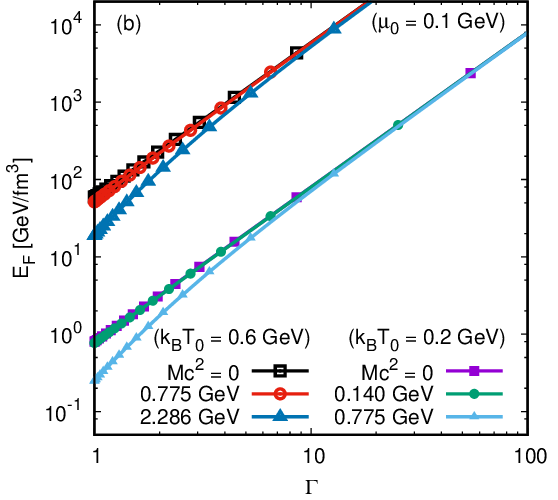} 
\end{tabular}
\caption{(a) Numerical results for the energy density 
$E_F$ \eqref{eq:RKT_gen_F}, in  ${\rm GeV} / {\rm fm}^3$ at $\mu_0 = 0.1\ {\rm GeV}$ 
and $k_{\rm {B}} T_0 = 0.2\ {\rm GeV}$, 
for $Mc^2 = 0$, $0.14\ {\rm GeV}$, 
$0.775\ {\rm GeV}$ and $1.116\ {\rm GeV}$.
(b) Log-log plot of $E_F$, at two temperatures ($k_{\rm {B}} T_0 = 0.2\ {\rm GeV}$ and 
$0.6\ {\rm GeV}$), for various masses.
The number of degrees of freedom was set to 
$g_{\rm {F}} = 6$.}
\label{fig:RKT_rho}
\end{figure}

In figure~\ref{fig:RKT_rho} we plot the radial profile of the energy density $E_{\rm {F}}$ (\ref{eq:RKT_gen_F}) as a function of $\rho $ (left-hand-plot, linear scale) and as a function of the Lorentz factor $\Gamma $ \eqref{eq:RKT_sol} (right-hand-plot, logarithmic scale). 
As expected, the energy diverges on the SLS for all values of the particle mass. 
The results for pions and massless particles are very nearly identical; for larger values of the mass the energy $E_{\rm {F}}$ is lower everywhere. 
However, close to the SLS the results for massive particles are indistinguishable from those for massless particles.
Similar behaviour is observed for the pressure $P_{\rm {F}}$ and
charge density $Q_{\rm {F}}$ \cite{Ambrus:2019ayb,Ambrus:2019khr}.
This is in agreement with the analytic work in the zero chemical potential case \cite{Ambrus:2015} (see also \cite{Florkowski:2013} for details of relevant techniques), where it was found that the ${\mathcal {O}}(M^{2})$ corrections due to the mass make subleading contributions 
as the SLS is approached. 

We now consider more closely the effect of varying the particle mass for both scalars and fermions.
To make the comparison relevant, we consider the energy 
density per particle degree of freedom, which amounts 
to dividing $E_{\rm F}$ by $2g_{\rm {F}}$ (the factor of two
is required since the particle and anti-particle 
states are explicitly taken into account) and 
$E_{\rm S}$ by $g_{\rm {S}}$. Furthermore, we consider 
the case of vanishing chemical potential, 
$\mu_0 =0$, since we have not introduced this 
quantity for scalars.

\begin{figure}
    \centering
\begin{tabular}{cc}
 \includegraphics[width=0.46\linewidth]{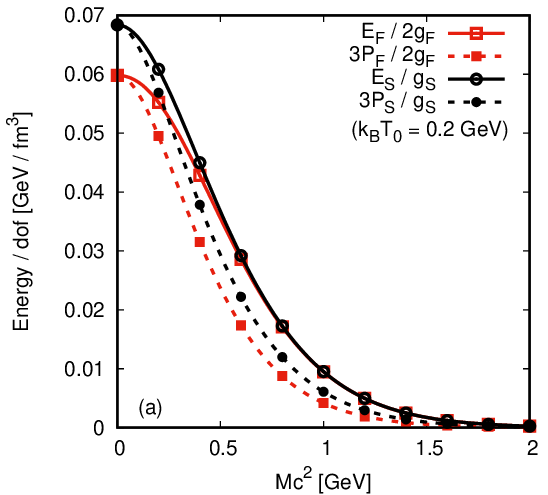} &
 \includegraphics[width=0.45\linewidth]{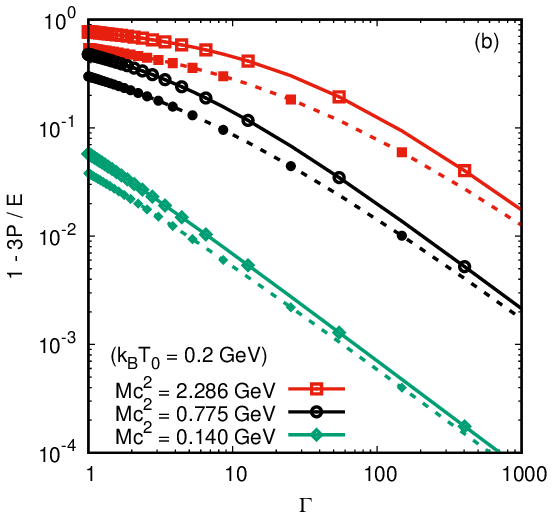} 
\end{tabular}
\caption{(a) The mass dependence of the energy (continuous 
lines and empty symbols) and pressure (dashed lines and 
filled symbols) per dof, 
computed for the Fermi-Dirac (squares) and 
Bose-Einstein (circle) statistics.
(b) The dependence on the Lorentz factor 
$\Gamma$ \eqref{eq:RKT_sol} of the
quantity $1 - 3P / E$, evaluated for 
the Fermi-Dirac (continuous lines and empty symbols)
and Bose-Einstein (dashed lines and filled symbols) 
statistics, for various values of the particle mass.}
\label{fig:RKT_mass}
\end{figure}

Figure \ref{fig:RKT_mass}(a) shows the effect of the particle  mass on the energy density and pressure on the rotation axis. For both scalars and fermions, these quantities decrease as the particle mass increases.
In figure \ref{fig:RKT_mass}(b) we plot the quantity $1 - 3 P / E$, which vanishes in the massless limit.
For a constant value of the Lorentz factor $\Gamma$,  we see that 
$1 - 3P / E$ increases as the mass is increased, thus the 
ratio $3P / E$ decreases. 
As the SLS is approached, $1 - 3P/E$ decreases, 
% as a power of $\Gamma^{-1}$,
showing that in the vicinity of the SLS, the gas behaves as though
its constituents were massless.

\section{Quantum rigidly-rotating thermal states}
\label{sec:therm}

We now consider the generalization from RKT to QFT, and examine how rigidly-rotating quantum states may be defined.
In the quantization process, the microscopic momenta are promoted to quantum operators.  
Thermal states at a temperature $T_{0}$ are defined such that the t.e.v. of an operator ${\widehat {A}}$ takes the form \cite{Kapusta:1989}:
\begin{equation}
 \braket{\widehat{A}}_{T_{0}} = Z^{-1} {\rm Tr} (\widehat{\rho} \widehat{A}),
 \label{eq:thermexp}
\end{equation}
where $Z={\rm {Tr}}{\widehat {\rho }}$ is the partition function and ${\widehat {\rho }}$ is the Boltzmann factor, which we define below.
The trace is performed over Fock space, that is, the space of all states of the quantum field containing $n$ particles (or anti-particles), for $n=0,1,2,\ldots $.

For a rigidly-rotating state with temperature $T_{0}$  on the axis of rotation,
the Boltzmann factor for a scalar field is given by \cite{Vilenkin:1980}:
\begin{equation}
 \widehat{\rho}_{\rm {S}} = \exp\left[-(\widehat{H}_{\rm {S}} - 
 \Omega \widehat{M}_{\rm {S}}^z ) /T_{0} \right],
 \label{eq:rhoS}   
\end{equation}
where ${\widehat {H}}_{\rm {S}}$ is the scalar Hamiltonian operator and  ${\widehat {M}}_{\rm {S}}^{z}$ is the $z$-component of the scalar angular momentum operator.
For a fermion field, we  include a chemical potential $\mu _{0}$ on the axis of rotation, which is conjugate to the charge operator. 
The Boltzmann factor for a fermion field is then given by \cite{Vilenkin:1980}:
\begin{equation}
 \widehat{\rho}_{\rm {F}} = \exp\left[-(\widehat{H}_{\rm {F}} - 
 \Omega \widehat{M}_{\rm {F}}^z - \mu _{0} \widehat{Q}_{\rm {F}}) /T_{0} \right],
 \label{eq:rhoF}
\end{equation}
where $\widehat{H}_{\rm {F}}$ is the fermion Hamiltonian operator, 
$\widehat{M}_{\rm {F}}^z$ is the $z$-component of the total 
fermion angular momentum operator and $\widehat{Q}_{\rm {F}}$ is the fermion charge operator.

In order to perform the trace over Fock space in (\ref{eq:thermexp}), 
we need to define particle creation and annihilation operators acting on the states. 
For a neutral scalar field, we denote the particle annihilation operators by ${\hat {a}}_{j}$, where $j$ labels the 
quantum properties
of the annihilated particle.
For a fermion field, the operators ${\hat {b}}_{j}$ annihilate fermions, while the ${\hat {d}}_{j}$ operators annihilate anti-fermions. 
In all cases, the adjoint operators are the corresponding particle creation operators.
For scalars, the particle creation and annihilation operators satisfy the canonical commutation relations
\begin{equation}
    [ {\hat {a}}_{j}, {\hat {a}}_{j'}^{\dagger } ] = {\hat {a}}_{j} {\hat {a}}_{j'}^{\dagger } - {\hat {a}}_{j'}^{\dagger }{\hat {a}}_{j} = \delta _{j,j'}, \qquad 
    [ {\hat {a}}_{j}, {\hat {a}}_{j'}] = 0 =
    [ {\hat {a}}_{j}^{\dagger }, {\hat {a}}_{j'}^{\dagger } ] ,
    \label{eq:commutation}
\end{equation}
where $\delta _{j,j'}$ vanishes unless the labels $j$ and $j'$ are identical.
For fermions, canonical anti-commutation relations hold, so that, for the particle operators:
\begin{equation}
    \{\hat{b}_j, \hat{b}^\dagger_{j'}\} = \hat{b}_j \hat{b}^\dagger_{j'} + 
\hat{b}^\dagger_{j'} \hat{b}_j = \delta_{j,j'}, \qquad
\{\hat{b}_j, \hat{b}_{j'}\} = 0 =\{\hat{b}_j^{\dagger  }, \hat{b}^\dagger_{j'}\},
\label{eq:anticommutation}
\end{equation}
and similar relations hold for the anti-particle operators.

Using the particle/anti-particle states corresponding to the above creation and annihilation operators, we consider a quantization which is compatible with the operator ${\widehat {\rho }}$, so that, for scalars:
\begin{equation}
    {\widehat {\rho }}_{\rm {S}} {\hat {a}}_{j}^{\dagger } \left( {\widehat {\rho }}_{\rm {S}} \right) ^{-1} = 
    e^{-\left( E_{j}-\Omega m_{j} \right)/T_{0}} {\hat {a}}_{j}^{\dagger },
    \label{eq:rhoarho}
\end{equation}
where $E_{j}$ is the energy of the created particle, and $m_{j}=0,\pm 1,
\pm 2,\ldots $ is the $z$-component of the angular momentum. 
Similarly, for fermions we assume that
\begin{equation}
 \widehat{\rho}_{\rm {F}} \hat{b}_j^\dagger \left( \widehat{\rho}_{\rm {F}} \right) ^{-1} = 
 e^{-\left(E_j - \Omega m_j - \mu _{0}\right) /T_{0}}
 \hat{b}_j^\dagger, \qquad
 \widehat{\rho}_{\rm {F}} \hat{d}_j^\dagger \left(\widehat{\rho}_{\rm {F}} \right) ^{-1} = 
 e^{-\left(E_j - \Omega m_j + \mu _{0} \right)/T_{0}}
 \hat{d}_j^\dagger,
 \label{eq:rhobrho}
\end{equation}
where $m_j = \pm \frac{1}{2}, \pm \frac{3}{2}, \dots$ is the projection 
of the total fermion angular momentum on the $z$-axis.
The quantities (\ref{eq:rhoarho}, \ref{eq:rhobrho}) depend on the energy ${\widetilde {E}}_{j}$  of the particle as seen by a co-rotating observer:
\begin{equation}
    {\widetilde {E}}_{j} = E_{j}- \Omega m_{j} .
    \label{eq:Etilde}
\end{equation}
Using the canonical 
commutation/anti-commutation
relations (\ref{eq:commutation}, \ref{eq:anticommutation}), together with (\ref{eq:rhoarho}, \ref{eq:rhobrho}), we find the t.e.v.s of the number operators for scalars to be \cite{Vilenkin:1980,Itzykson:1980rh}:
\begin{equation}
    \braket{ {\hat {a}}^{\dagger }_{j}{\hat {a}}_{j'} }_{T_{0}}
    = \frac{\delta _{j,j'}}{\exp [ {\widetilde {E}}_{j}/T_{0} ] - 1},
    \label{eq:ablocks}
\end{equation}
while for fermions we have
\begin{equation}
 \braket{ \hat{b}^\dagger_j \hat{b}_{j'} }_{T_{0}} = 
 \frac{\delta_{j,j'}}
 {\exp[(\widetilde{E}_j - \mu _{0})/T_{0}] + 1}, \qquad
 \braket{ \hat{d}^\dagger_j \hat{d}_{j'} }_{T_{0}} = \frac{\delta_{j,j'}}
 {\exp[(\widetilde{E}_j + \mu _{0})/T_{0}] + 1}. 
 \label{eq:bblocks}
\end{equation}
The t.e.v.s (\ref{eq:ablocks}, \ref{eq:bblocks}) have the expected Bose-Einstein/Fermi-Dirac thermal distributions in terms of the co-rotating energy ${\widetilde {E}}_{j}$.

Consider first the scalar field t.e.v.~(\ref{eq:ablocks}). 
This has the correct zero-temperature limit only if ${\widetilde {E}}_{j}>0$. 
Even with this restriction, it can be seen that (\ref{eq:ablocks}) diverges when ${\widetilde {E}}_{j}\rightarrow 0$, leading to the divergence of all t.e.v.s (\ref{eq:thermexp})  \cite{Vilenkin:1980,Duffy:2002ss}.
From this we deduce that rigidly-rotating thermal states cannot be defined for a quantum scalar field on unbounded Minkowski space-time \cite{Vilenkin:1980,Duffy:2002ss,Ambrus:2014uqa}.

To understand this result, we consider how a quantum vacuum state is defined for a scalar field. 
In the canonical quantization approach to QFT, one starts with an orthonormal basis of scalar field modes $\phi _{j}$ which are solutions of the Klein-Gordon equation for a massive scalar field,
$\left( \partial _{\mu }\partial ^{\mu } + M^{2}\right) \phi _{j} =0$.
The scalar field operator ${\widehat {\Phi }}$ is then written as a sum over these field modes and their complex conjugates
\begin{equation}
    {\widehat {\Phi }} = \sum _{j} \left[ {\hat {a}}_{j}\phi _{j} + {\hat {a}}_{j}^{\dagger } \phi _{j}^{*} \right] ,
    \label{eq:phiexpansion}
\end{equation}
where the expansion coefficients are the particle creation and annihilation operators.
In order that the creation and annihilation operators satisfy the canonical commutation relations (\ref{eq:commutation}), it must be the case that
\begin{equation}
    \braket{\phi _{j}, \phi _{j'}} = \delta _{j,j'}, \qquad 
    \braket{\phi ^{*}_{j}, \phi _{j'}^{*} } = -\delta _{j,j'}, \qquad
    \braket{ \phi _{j}, \phi ^{*}_{j'}} = 0,
    \label{eq:KGinner}
\end{equation}
where $\braket{\, , \,}$ is the Klein-Gordon inner product, defined for two solutions $\phi _{j}$, 
$\phi_{j'}$
of the Klein-Gordon equation  by the following integral over a constant-$t$ surface:
\begin{equation}
    \braket{\phi_{j}, \phi _{j'}}= i\int d^{3}x \left( \phi _{j}^{*} \partial ^{t} \phi _{j'} - \phi _{j'} \partial ^{t}\phi _{j}^{*} \right) .
\end{equation}
In particular, the modes $\phi _{j}$ corresponding to particles must have positive norm $\braket{\phi _{j}, \phi _{j}}$, while those modes $\phi _{j}^{*}$ corresponding to antiparticle modes must have negative norm.
This restricts whether modes can be labelled as ``particle'' or ``antiparticle''.  
Calculating the inner product for a particle mode with energy $E_{j}$, we find
\begin{equation}
   \braket{\phi _{j}, \phi _{j'}} = \frac{E_{j}}{\left| E_{j} \right| }\delta _{j,j'}, 
\end{equation}
and hence the relations (\ref{eq:KGinner}) hold only if the energy $E_{j}$ of the mode $\phi _{j}$ is positive, $E_{j}>0$ \cite{Letaw:1979wy}.
The vacuum state  $|0\rangle $ is then defined as that state which is annihilated by the particle annihilation operators,
${\hat {a}}_{j}|0 \rangle = 0$,
and is simply the (stationary) Minkowski vacuum.
For a quantum scalar field, it is not possible to make the choice  ${\widetilde {E}}_{j}>0$ because, for fixed ${\widetilde {E}}_{j}>0$, there will be modes with sufficiently large and negative $m_{j}$ for which
$E_{j}={\widetilde {E}}_{j}+\Omega m_{j}<0$, so that (\ref{eq:KGinner}) no longer holds and we do not have a valid quantization \cite{Letaw:1979wy}.  
Since there is no rotating vacuum for a quantum scalar field, rotating thermal states for a quantum scalar field are also ill-defined.

One resolution of this difficulty is to insert a reflecting boundary inside the SLS \cite{Vilenkin:1980,Duffy:2002ss}.
The presence of the boundary means that the energy $E_{j}$ of the scalar field modes is quantized, and, if the boundary is inside the SLS, it can be shown that ${\widetilde {E}}_{j}>0$ for all $m_{j}$ \cite{Duffy:2002ss,Nicolaevici:2001}. 
In this case a rotating vacuum state (and also rotating thermal states) can be defined for a quantum scalar field \cite{Duffy:2002ss}.

In view of these difficulties for a quantum scalar field, for the rest of this chapter we restrict our attention to a quantum fermion field on unbounded Minkowski space-time. 
First we consider whether a rotating vacuum state can be defined in canonical quantization.
Beginning with an orthonormal basis of particle mode solutions $U_{j}$  and anti-particle mode solutions $V_{j}$ of the Dirac equation (which will be discussed in more detail in the next section), the fermion field operator is written as
\begin{equation}
    {\widehat {\Psi }} = \sum _{j} \left[ {\hat {b}}_{j} U _{j} + {\hat {d}}_{j}^{\dagger } V_{j}  \right] ,
    \label{eq:psisumj}
\end{equation}
where the operators ${\hat {b}}_{j}$ and ${\hat {d}}_{j}$ satisfy the canonical anti-commutation relations (\ref{eq:anticommutation}).
In contrast to the scalar field case, all particle and antiparticle modes $U_{j}$, $V_{j}$ have positive Dirac norm, 
resulting in a greater freedom to label modes as ``particle'' or ``anti-particle''.  
This in turn leads to a greater freedom in how vacuum states (and therefore also thermal states) are defined \cite{Ambrus:2014uqa}.

One possible quantization is to define ``particle'' modes as having positive energy $E_{j}$ \cite{Vilenkin:1980}.
As in the scalar case, the resulting vacuum is simply the usual (nonrotating) Minkowski vacuum state.
However, for fermions there is another possibility \cite{Iyer:1982ah}: 
particle modes can be defined by setting ${\widetilde {E}}_{j}>0$. 
This leads to a well-defined quantization and a rotating vacuum state.  
Furthermore, with this definition the t.e.v.s (\ref{eq:bblocks}) have the correct zero-temperature limit, with 
contributions only from modes below the Fermi level,  for 
which $\widetilde{E}_j < \mu _{0}$ for particle modes and 
${\widetilde {E}}_{j} < -\mu _{0}$ 
for antiparticle modes (we remind the reader that $\widetilde{E}_j > 0$ always holds when the rotating vacuum is employed). 
This is in agreement with the corresponding result \eqref{eq:RKT_EF} in the RKT case, and is sufficient to ensure that 
there are no temperature- and chemical potential-independent 
contributions to t.e.v.s.

\section{Mode solutions in cylindrical coordinates}\label{sec:modes}

Our purpose for the remainder of this chapter is to compute t.e.v.s of observables for a quantum fermion field of mass $M$, and compare the results with those for the RKT approach in section \ref{sec:RKT:macro}.
In this section we lay the groundwork for our computation by considering in more detail the fermion mode solutions discussed schematically in the previous section.
Since we are interested in rigidly-rotating states, we work in cylindrical coordinates $x^\mu = (t, \rho, \varphi, z)$ and follow the approach of \cite{Ambrus:2014uqa,Ambrus:2015lfr}.

The evolution of a free Dirac field with mass $M$ is governed by the
least-action principle, starting from the action:
\begin{equation}
 S_{\rm {F}} = i \int d^4x \, \mathcal{L}, \qquad 
 \mathcal{L} = \frac{i}{2} 
 \left(\overline{\psi} \slashed{\partial} \psi - 
 \overline{\slashed{\partial} \psi} \psi\right) - 
 M \overline{\psi} \psi,
 \label{eq:action}
\end{equation}
where the Feynman slash denotes contraction with the gamma 
matrices $\slashed{\partial} = \gamma^\mu \partial_\mu$.
The gamma matrices satisfy the canonical anti-commutation relations 
$\{\gamma^\mu, \gamma^\nu\} = 2 \eta ^{\mu\nu}$ 
and in this chapter, we work with the Dirac representation:
\begin{equation}
 \gamma^t = 
 \begin{pmatrix}
  1 & 0 \\
  0 & -1
 \end{pmatrix}, \qquad \gamma^i =
 \begin{pmatrix}
  0 & \sigma^i\\
  -\sigma^i & 0
 \end{pmatrix},
 \label{eq:gamma}
\end{equation}
where the Pauli matrices are given by:
\begin{equation}
 \sigma^x = 
 \begin{pmatrix}
  0 & 1 \\ 
  1 & 0
 \end{pmatrix}, \qquad \sigma^y = 
 \begin{pmatrix}
  0 & -i \\ 
  i & 0
 \end{pmatrix}, \qquad \sigma^z =
 \begin{pmatrix}
  1 & 0 \\
  0 & -1
 \end{pmatrix}.
\end{equation}
We are considering four-spinors $\psi $, which have Dirac adjoint ${\overline {\psi }} =  \psi ^{\dagger }\gamma ^{t}$.
Demanding that the variation of the action $S_{\rm {F}}$ (\ref{eq:action}) with respect to  the $\overline{\psi}$ degree of freedom vanishes yields the 
Dirac equation
\begin{equation}
 (i\slashed{\partial} - M) \psi = 0.
 \label{eq:dirac}
\end{equation}

As outlined in the previous section, in order to construct t.e.v.s we first require a set of particle modes $\{U_j\}$ and anti-particle modes $\{V_j\}$ satisfying 
the Dirac equation \eqref{eq:dirac}. 
Given a particle mode $U_{j}$, the corresponding anti-particle mode $V_j$ 
is related to $U_{j}$ by the charge conjugation 
operation:
\begin{equation}
 V_j = i \gamma^y U_j^*.\label{eq:cc}
\end{equation}
In deriving the formal expressions for rigidly-rotating t.e.v.s in section~\ref{sec:therm}, we have assumed a quantization compatible with ${\widehat {\rho }}$, see \eqref{eq:rhobrho}.
This requires that the following 
commutation relations must hold:
\begin{align}
 [\widehat{H}_{\rm {F}}, \hat{b}_j^\dagger] =& E_j \hat{b}_j^\dagger, &
 [\widehat{M}^z_{\rm {F}}, \hat{b}_j^\dagger] =& m_j \hat{b}_j^\dagger, &
 [\widehat{Q}_{\rm {F}}, \hat{b}_j^\dagger] =& \hat{b}_j^\dagger, \nonumber\\
 [\widehat{H}_{\rm {F}}, \hat{d}_j^\dagger] =& E_j \hat{d}_j^\dagger, &
 [\widehat{M}^z_{\rm {F}}, \hat{d}_j^\dagger] =& m_j \hat{d}_j^\dagger, &
 [\widehat{Q}_{\rm {F}}, \hat{d}_j^\dagger] =& -\hat{d}_j^\dagger.
\end{align}
Taking into account the expression for the conserved operators 
in the classical Dirac field theory, 
\begin{equation}
 H_{\rm {F}} = i \partial_t, \qquad 
 M^z_{\rm {F}} = -i\partial_\varphi + S^z, 
\end{equation}
where the $z$-projection of the spin operator $S^z$ is given by:
\begin{equation}
 S^z = \frac{1}{2}
 \begin{pmatrix}
  \sigma^z & 0 \\ 
  0 & \sigma^z 
 \end{pmatrix},
\end{equation}
the particle mode solutions $U_j$ must thus be chosen to be simultaneous eigenfunctions 
of $H_{\rm {F}}$ and $M^z_{\rm {F}}$:
\begin{equation}
 H_{\rm {F}} U_j = E_j U_j, \qquad 
 M^z_{\rm {F}} U_j = m_j U_j.
\end{equation}
The above eigenvalue equations are insufficient to specify the particle mode solutions 
uniquely. The remaining degrees of freedom can be fixed by 
choosing $U_j$ to be eigenfunctions of the longitudinal momentum 
operator $P^z_{\rm {F}} = -i\partial_z$ and of the helicity operator 
$W_0 = \bm{J} \cdot \bm{P}_{\rm {F}} / 2p$ (where $p$ is the magnitude of the momentum):
\begin{equation}
 P^z_{\rm {F}} U_j = k_j U_j, \qquad 
 W_0 U_j = \lambda_j U_j,
\end{equation}
where $k_{j}$ and $\lambda _{j}$ are real constants.
The expression for $W_0$ can be obtained as follows:
\begin{equation}
 W_0 = 
 \begin{pmatrix}
  h & 0 \\
  0 & h
 \end{pmatrix}, \qquad 
 h = \frac{\bm{\sigma} \cdot \bm{P}_{\rm {F}}}{2p} = 
 \frac{1}{2p}
 \begin{pmatrix}
  P^z_{\rm {F}} & P_- \\
  P_+ & -P^z_{\rm {F}},
 \end{pmatrix},
\end{equation}
where $\bm{P}_{\rm F} = -i \nabla$, while
$P_\pm$ are defined in terms of cylindrical coordinates as:
\begin{equation}
 P_\pm = P^x_{\rm {F}} \pm iP^y_{\rm {F}} = -i e^{\pm i\varphi} 
 (\partial_\rho \pm i \rho^{-1} \partial_\varphi).
 \label{eq:Ppm}
\end{equation}
It can be shown that $W_0^2 = \frac{1}{4}$. The eigenvalues 
$\lambda_j = 1/2$ and $-1/2$ correspond to positive and 
negative helicity, respectively.

The Dirac equation \eqref{eq:dirac} can 
be written with respect to the above operators as:
\begin{equation}
 \begin{pmatrix}
  H_{\rm {F}} - M & -2ph \\
  2ph & -H_{\rm {F}} -M
 \end{pmatrix} \psi = 0.
 \label{eq:dirac_op}
\end{equation}
The operators $H_{\rm {F}}$ and $P^z_{\rm {F}}$ are diagonal with 
respect to the spinor structure, thus the corresponding 
eigenvalue equations can be solved immediately:
\begin{equation}
 U_j = \frac{{\mathcal{K}}_j}{2\pi} e^{-i E_j t + i k_j z} u_j, \qquad 
 u_j = 
 \begin{pmatrix}
  \mathcal{C}_j^{-} \phi_j \\ 
  \mathcal{C}_j^{+} \phi_j,
 \end{pmatrix},
 \label{eq:Ujsep}
\end{equation}
where $u_j$ is a four-spinor which depends only on $\varphi$ and $\rho$ and ${\mathcal {K}}_{j}$ is a normalization constant.
In \eqref{eq:Ujsep}, $\mathcal{C}_j^{\pm}$ are integration constants and
$\phi_j$ is a two-spinor satisfying the remaining two 
eigenvalue equations, namely:
\begin{equation}
 \left(-i \partial_\varphi + \frac{1}{2} \sigma^z\right) \phi_j =
 m_j \phi_j, \qquad h \phi_j = \lambda_j \phi_j.
 \label{eq:eigen_phi}
\end{equation}
Substituting (\ref{eq:Ujsep}) into the Dirac equation \eqref{eq:dirac_op} gives
\begin{equation}
 \begin{pmatrix}
  E_j - M & -2p_j \lambda_j \\
  2p_j \lambda_j & -E_j - M
 \end{pmatrix} u_j = 0,
\end{equation}
where the magnitude of the momentum is now $p_{j}$, and from this the following relation can be established for $\mathcal{C}_j^{\pm}$:
\begin{equation}
 \mathcal{C}_j^- = \frac{2 \lambda_j p_j}{E_j - M} \mathcal{C}_j^+.
\end{equation}

Next we consider the angular momentum equation, the first relation in  \eqref{eq:eigen_phi}, which allows $\phi_j$ to be written in the form:
\begin{equation}
 \phi_j = 
 \begin{pmatrix}
  \phi_j^- e^{i (m_j - \frac{1}{2}) \varphi} \\
  \phi_j^+ e^{i (m_j + \frac{1}{2}) \varphi}
 \end{pmatrix},
\end{equation}
where $m_j = \pm \frac{1}{2}, \pm \frac{3}{2}, \dots$ is an
odd half-integer, while $\phi_j^\pm \equiv \phi_j^\pm(\rho)$ 
are functions which depend only on the radial coordinate $\rho$. 
Taking into account the result [from \eqref{eq:Ppm}] $P_+ P_- = P_- P_+ = -\partial_\rho^2 - 
\rho^{-1} \partial_\rho - \rho^{-2} \partial_\varphi^2$,
the second relation in (\ref{eq:eigen_phi}) reduces to:
\begin{equation}
 \left[\rho^2 \frac{\partial^2}{\partial \rho^2} + 
 \rho \frac{\partial}{\partial \rho} + q_j^2 \rho^2 - 
 \left(m_j \pm \frac{1}{2}\right)^2\right] 
 \phi_j^{\pm} = 0,\label{eq:phijpm_eq}
\end{equation}
where the longitudinal momentum $q_j$ is defined by
\begin{equation}
 q_j = \sqrt{p_j^2 - k_j^2} = \sqrt{E_j^2 - k_j^2 - M^2}.
\end{equation}
Equation \eqref{eq:phijpm_eq} can readily be identified with 
the Bessel equation \cite{Olver:2010}, having  two linearly independent 
solutions $J_{m \pm 1/2}(q\rho)$ 
and $Y_{m \pm 1/2}(q \rho)$. Demanding regularity at the 
origin discards the Neumann function $Y_{m \pm 1/2}(q \rho)$, and therefore
\begin{equation}
\phi^\pm_j = \mathcal{N}_j^\pm J_{m_j \pm \frac{1}{2}}(q_j \rho).
\end{equation}
The connection between the integration constants 
$\mathcal{N}_j^+$ and $\mathcal{N}_j^-$ can be established 
by noting that the operators $P_\pm$ act as ladder operators, 
in the sense that:
\begin{equation}
 P_\pm e^{i(m_j \mp \frac{1}{2}) \varphi} 
 J_{m_j \mp \frac{1}{2}}(q_j \rho) = 
 \pm i q_j e^{i (m_j \pm \frac{1}{2}) \varphi}
 J_{m_j \pm \frac{1}{2}}(q_j \rho),
\end{equation}
where the following properties were employed \cite{Olver:2010}:
\begin{align}
 J'_{m_j+\frac{1}{2}}(q_j \rho) =& 
 J_{m_j-\frac{1}{2}}(q_j \rho) - 
 \frac{m_j + \frac{1}{2}}{q_j \rho} 
 J_{m_j + \frac{1}{2}}(q_j \rho), \nonumber\\
 J'_{m_j-\frac{1}{2}}(q_j \rho) =& 
 -J_{m_j+\frac{1}{2}}(q_j \rho) + 
 \frac{m_j - \frac{1}{2}}{q_j \rho} 
 J_{m_j - \frac{1}{2}}(q_j \rho).
 \label{eq:Jprime}
\end{align}
The helicity equation [the second relation in \eqref{eq:eigen_phi}] then yields
\begin{equation}
 {\mathcal {N}}_{j}^{+} = \frac{iq_{j}}{k_{j}+2p_{j}\lambda _{j}} {\mathcal {N}}_{j}^{-} = 
 2i\lambda_j \frac{\mathfrak{p}_j^-}{\mathfrak{p}_j^+}
 {\mathcal{N}}_{j}^{-}, 
 \end{equation}
 with
 \begin{equation}
 \mathfrak{p}_j^\pm = 
 \left(1 \pm \frac{2\lambda_j k_j}{p_j}\right)^{1/2}.
\end{equation}
Noting that an overall normalization constant, 
$\mathcal{N}_j^- \sqrt{2} / \mathfrak{p}_j^+$, can be 
absorbed into ${\mathcal {K}}_{j}$ in \eqref{eq:Ujsep},
we write $\phi_j$ in the form:
\begin{equation}
 \phi_j = \frac{1}{\sqrt{2}} 
 \begin{pmatrix}
  \mathfrak{p}^+_j e^{i (m_j - \frac{1}{2}) \varphi} 
  J_{m_j - \frac{1}{2}} (q_j \rho) \\
  2i\lambda_j \mathfrak{p}^-_j e^{i (m_j + \frac{1}{2}) \varphi} 
  J_{m_j + \frac{1}{2}} (q_j \rho) \\
 \end{pmatrix}.
 \label{eq:phij}
\end{equation}
Introducing the angle $\vartheta_j$ made by the momentum vector with 
the $z$-direction, so that $k_j = p_j \cos\vartheta_j$ 
(with $0 \le \vartheta_j \le \pi$), it can be seen that 
\begin{equation}
 \frac{1}{\sqrt{2}} \mathfrak{p}_j^\pm = 
 \left(\frac{1}{2} \pm \lambda_{j} \right) \cos\frac{\vartheta_j}{2} +
 \left(\frac{1}{2} \mp \lambda_{j} \right) \sin\frac{\vartheta_j}{2}.
\end{equation}
Thus, the two-spinor $\phi_j$ can be written compactly as follows (where we have explicitly written out all the parameters on which this depends):
\begin{equation}
 \phi_{p,k,m}^{1/2} = 
 \begin{pmatrix}
  \cos\frac{\vartheta}{2} e^{i (m - \frac{1}{2}) \varphi} 
  J_{m - \frac{1}{2}} (q \rho) \\
  i\sin\frac{\vartheta}{2} e^{i (m + \frac{1}{2}) \varphi} 
  J_{m + \frac{1}{2}} (q \rho)
 \end{pmatrix}, \qquad
 \phi_{p,k,m}^{-1/2} = 
 \begin{pmatrix}
  \sin\frac{\vartheta}{2} e^{i (m - \frac{1}{2}) \varphi} 
  J_{m - \frac{1}{2}} (q \rho) \\
  -i\cos\frac{\vartheta}{2} e^{i (m + \frac{1}{2}) \varphi} 
  J_{m + \frac{1}{2}} (q \rho)
 \end{pmatrix}.
 \label{eq:phij2}
\end{equation}
Using the identity 
\begin{equation}
 \sum_{n = -\infty}^\infty J_{n}^2(q_j \rho) = 1,
 \label{eq:Jsum}
\end{equation}
where the sum runs over all integers $n \in \mathbb{Z}$, it can be established that the two-spinors $\phi _{j}$ \eqref{eq:phij2} satisfy the normalization condition
\begin{equation}
 \sum_{m = -\infty}^\infty \phi_{p,k,m}^{\lambda,\dagger} \phi_{p,k,m}^{\lambda'} 
 = \delta_{\lambda,\lambda'},\label{eq:phinorm}
\end{equation}

We now return to the four-spinors $u_{j}$ \eqref{eq:Ujsep},
for which
we impose the normalization condition
\begin{equation}
 \sum_{m = -\infty}^\infty u_{E,k,m}^{\lambda,\dagger} u_{E,k,m}^{\lambda'}
 = \delta_{\lambda,\lambda'}.
 \label{eq:unorm}
\end{equation}
This can be achieved by setting 
$\mathcal{C}_j^+ = (2\lambda_j E_j / |E_j|)
\mathfrak{E}_j^- / \sqrt{2}$, such that
\begin{equation}
 u_j = \frac{1}{\sqrt{2}} 
 \begin{pmatrix}
  \mathfrak{E}^+_j \phi_j \\ 
  \frac{2 \lambda_j E_j}{|E_j|} \mathfrak{E}^-_j \phi_j
 \end{pmatrix}, \qquad 
 \mathfrak{E}^\pm_j = \left(1 \pm \frac{M}{E_j}\right)^{1/2},
\end{equation}
where $E_j / |E_j|$ is the sign of $E_j$.

The final piece of the puzzle is to establish unit norm for the modes $U_j$ \eqref{eq:Ujsep}.
This is achieved using the Dirac inner product, defined for two solutions $\psi$ and $\chi$ of the Dirac equation \eqref{eq:dirac} by:
\begin{equation}
 \braket{\psi, \chi} = \int d^3x\, \overline{\psi} \gamma^t \chi ,
\end{equation}
where the integration is taken over a constant-$t$ surface.
Performing the integral with respect to cylindrical coordinates and 
using the relation
\begin{equation}
 \int_0^\infty d\rho\, \rho\, J_{m + \frac{1}{2}}(q_j \rho) 
 J_{m + \frac{1}{2}}(q_{j'} \rho) = 
 \frac{\delta(q_j - q_{j'})}{q_j},
\end{equation}
it can be seen that, with ${\mathcal {K}}_{j}=1$, we have the required normalisation condition
\begin{align}
 \braket{U_j, U_{j'}} =& \delta_{\lambda_j, \lambda_{j'}} 
 \delta_{m_j, m_{j'}} \delta(k_j - k_{j'}) 
 \frac{\delta(q_j - q_{j'})}{q_j} \theta(E_j E_{j'}) \nonumber\\
 =& \delta_{\lambda_j, \lambda_{j'}} 
 \delta_{m_j, m_{j'}} \delta(k_j - k_{j'}) 
 \frac{\delta(E_j - E_{j'})}{|E_j|}.
 \label{eq:normalization}
\end{align}

We therefore write the particle modes $U_{j}$ as
\begin{equation}
 U_{E,k,m}^\lambda = \frac{e^{-i E t + i k z}}{2\pi} u_{E,k,m}^\lambda, \qquad 
 u_{E,k,m}^\lambda = \frac{1}{\sqrt{2}} 
 \begin{pmatrix}
  \mathfrak{E}^+ \phi_{p,k,m}^\lambda \\
  \frac{2\lambda E}{|E|} 
  \mathfrak{E}^- \phi_{p,k,m}^\lambda
 \end{pmatrix}.
 \label{eq:Ujfinal}
\end{equation}
The four-spinors  $V_{j}$ corresponding to the anti-particle modes 
are then obtained via the charge conjugation operation \eqref{eq:cc}:
\begin{equation}
 V_{E,k,m}^\lambda = \frac{e^{i E t - i k z}}{2\pi} v_{E,k,m}^\lambda, \qquad 
v_{E,k,m}^\lambda = \frac{(-1)^{m-\frac{1}{2}}}{\sqrt{2}} 
\frac{i E}{|E|}
 \begin{pmatrix}
  \mathfrak{E}^- \phi_{p,-k,-m}^\lambda \\
  -\frac{2\lambda E}{|E|} 
  \mathfrak{E}^+ \phi_{p,-k,-m}^\lambda
 \end{pmatrix}.
 \label{eq:Vjfinal}
\end{equation}
The two-spinor $\phi_{p,k,m}^\lambda$ is defined in 
\eqref{eq:phij}, and also in \eqref{eq:phij2} 
in terms of the angle $\vartheta$ between the momentum vector 
and the $z$-axis.
Due to the relationship (\ref{eq:cc}) between the particle and anti-particle modes, the anti-particle modes $V_{j}$ also satisfy the normalization condition (\ref{eq:normalization}).
In particular, anti-particle modes, like particle modes, have positive Dirac norm. 
As discussed in the previous section, this is crucial for the definition of rigidly-rotating quantum states for fermions.

\section{Quantum stationary thermal expectation values} 
\label{sec:tevs0}

With a complete orthonormal basis of fermion modes constructed in the previous section, we are now in a position to compute t.e.v.s of physical quantities.
While our primary interest is in rigidly-rotating states, we first study the t.e.v.s for stationary, nonrotating states with vanishing angular speed $\Omega $.

At the level of the 
classical field theory, the CC $J^\mu$ and SET $T^{\mu\nu}$ can be constructed using Noether's 
theorem \cite{Itzykson:1980rh}:
\begin{equation}
 J^\mu = \overline{\psi} \gamma^\mu \psi, \qquad 
 T_{\mu\nu} = \frac{i}{2} 
 \left[\overline{\psi} \gamma_{(\mu} \partial_{\nu)} \psi - 
 \partial_{(\mu} \overline{\psi} \gamma_{\nu)} \psi\right].
 \label{eq:JT_aux}
\end{equation}
The trace of the SET 
is proportional to the FC $\overline{\psi }\psi $:
\begin{equation}
 T^\mu{}_\mu =  M \overline{\psi}\psi.
 \label{eq:FC_def}
\end{equation}
The generalisation to QFT is made by 
replacing the classical field $\psi$ with the corresponding 
quantum operator, $\widehat{\Psi}$. 
Due to the anti-commutation 
relations \eqref{eq:anticommutation} satisfied by the quantum operators, there is an ambiguity in the ordering of the action of the quantum 
operators on the Fock space states. 
For operators which are quadratic in the field operators,
such as those arising from (\ref{eq:JT_aux}, \ref{eq:FC_def}), 
and since we are working on flat space-time, 
this ambiguity can be overcome by introducing normal ordering, 
a procedure by which the vacuum expectation value (v.e.v.) is 
subtracted from the operator itself. 
For an operator $\widehat{A}$, the normal-ordered operator $:\widehat{A}:$ is therefore defined to be
\begin{equation}
 :\widehat{A}: = \widehat{A} - \braket{0|\widehat{A}|0}.
\end{equation}

Inserting the schematic mode expansion \eqref{eq:psisumj} in 
\eqref{eq:JT_aux}, the following expressions 
are obtained 
\begin{align}
:\widehat{\overline{\Psi}}\widehat{\Psi}: =& \sum_{j,j'} \left[
 \hat{b}_j^\dagger \hat{b}_{j'} \overline{U}_j U_{j'} - 
 \hat{d}_{j'}^\dagger \hat{d}_{j} \overline{V}_j V_{j'}\right],
 \nonumber \\
 :\widehat{J}^\mu: =& \sum_{j,j'} \left[ \hat{b}_j^\dagger \hat{b}_{j'} 
 \mathfrak{J}^\mu(U_j, U_{j'}) - \hat{d}_{j'}^\dagger \hat{d}_{j} 
 \mathfrak{J}^\mu(V_j, V_{j'}) \right],\nonumber\\
 :\widehat{T}_{\mu\nu}: =& \sum_{j,j'} \left[ \hat{b}_j^\dagger \hat{b}_{j'} 
 \mathcal{T}_{\mu\nu}(U_j, U_{j'}) - \hat{d}_{j'}^\dagger \hat{d}_{j} 
 \mathcal{T}_{\mu\nu}(V_j, V_{j'})\right], 
 \label{eq:schematic}
\end{align}
where we have introduced the sesquilinear forms $\mathfrak{J}^\mu(\psi,\chi)$ and 
$\mathcal{T}_{\mu\nu}(\psi,\chi)$  for notational 
brevity, based on the classical quantities \eqref{eq:JT_aux}:
\begin{equation}
 \mathfrak{J}^\mu(\psi,\chi) = \overline{\psi} \gamma^\mu \chi, \qquad 
 \mathcal{T}_{\mu\nu}(\psi,\chi) = \frac{i}{2} 
 \left[\overline{\psi} \gamma_{(\mu} \partial_{\nu)} \chi - 
 \partial_{(\mu} \overline{\psi} \gamma_{\nu)} \chi\right].
\end{equation}

As discussed in section \ref{sec:therm}, the nonrotating Minkowski vacuum is defined by taking 
all modes corresponding to 
the positive eigenvalues of the Hamiltonian ($E_j > 0$) as 
particle modes. This leads to the following decomposition of the 
field operator:
\begin{equation}
 \widehat{\Psi} = \sum_{\lambda = \pm \frac{1}{2}} 
 \sum_{m = -\infty}^\infty \int_M^\infty dE \, E\,
 \int_{-p}^p dk 
 \left[\hat{b}_{E,k,m}^\lambda U_{E,k,m}^\lambda + 
 \hat{d}_{E,k,m}^\lambda{}^\dagger V_{E,k,m}^\lambda\right], 
 \label{eq:vacs}
\end{equation}
where the spinor modes are given by (\ref{eq:Ujfinal}, \ref{eq:Vjfinal}).
Substituting the mode expansion \eqref{eq:vacs} into 
\eqref{eq:schematic}, and
using the relations \eqref{eq:bblocks}, we find the following t.e.v.s for a stationary (nonrotating) state at temperature $T_{0}$:
\begin{align}
 \braket{:\widehat{\overline{\Psi}} \widehat{\Psi}:}_{T_{0}} =& 
 \sum_{j} \left\{\frac{
 \overline{U}_j U_{j}}{\exp[(E_j - \mu _{0})/T_{0}] + 1} - 
 \frac{\overline{V}_j V_{j}}{\exp[(E_j + \mu _{0})/T_{0}] + 1}
 \right\}, \nonumber\\
 \braket{:\widehat{J}^\mu:}_{T_{0}} =& \sum_{j} \left\{
 \frac{\mathfrak{J}^\mu(U_j, U_{j})}
 {\exp[(E_j - \mu _{0})/T_{0}] + 1} - 
 \frac{\mathfrak{J}^\mu(V_j, V_{j})}
 {\exp[(E_j + \mu_0)/T_{0}] + 1}
 \right\},\nonumber\\
 \braket{:\widehat{T}_{\mu\nu}:}_{T_{0}} =& \sum_{j} \left\{
\frac{\mathcal{T}_{\mu\nu}(U_j, U_{j})}
{\exp[(E_j - \mu _{0})/T_{0}] + 1} - 
 \frac{\mathcal{T}_{\mu\nu}(V_j, V_{j})}
 {\exp[(E_j + \mu _{0})/T_{0}] + 1}  
 \right\}, \label{eq:tevs0_aux}
\end{align}
where $\widetilde{E}_j = E_j$ in the case when $\Omega = 0$.

\subsection{Fermion condensate}
\label{sec:tevs0:ppsi}

Using the charge conjugation property \eqref{eq:cc}, it can be shown that:
\begin{equation}
 \overline{V}_j V_j = \overline{U}_j^* \gamma^y \gamma^y U_j^* 
 = -(\overline{U}_j U_j)^*,
 \label{eq:VV}
\end{equation}
since $(\gamma^y)^2 = -1$.
Using the spinor mode \eqref{eq:Ujfinal},
we have
\begin{equation}
 \overline{U}_j U_j = \frac{M}{8\pi^2 E_j} 
 \left[J_{m_j}^+(q_j \rho) + \frac{2\lambda_j k_j}{p_j} 
 J_{m_j}^-(q_j \rho)\right],
 \label{eq:UU}
\end{equation}
where we define (we will need $J_m^\times(q\rho)$ later)
\begin{equation}
 J_m^\pm(q\rho) = J_{m - \frac{1}{2}}^2(q\rho) \pm
 J_{m + \frac{1}{2}}^2(q\rho), \qquad 
 J_m^\times(q\rho) = 2 J_{m - \frac{1}{2}}(q\rho) 
 J_{m + \frac{1}{2}}(q\rho).
 \label{eq:Jstar_def}
\end{equation}
Since $\overline{U}_j U_j$ is a real scalar, 
it can be seen that $\overline{V}_j V_j = -\overline{U}_j U_j$.
Furthermore, noting that the term proportional to 
$\lambda_{j}$ in \eqref{eq:UU} makes a vanishing contribution 
under the summation with respect to $\lambda _{j}$, 
the t.e.v.~of the FC, given in 
the first line of \eqref{eq:tevs0_aux}, is:
\begin{equation}
 \braket{:\widehat{\overline{\Psi}}\widehat{\Psi} :}_{T_{0}} = 
 \frac{M}{4\pi^2} \sum_{m = -\infty}^\infty \int_M^\infty dE 
 \left[\frac{1}{e^{(E - \mu _{0})/T_{0}} + 1} +
 \frac{1}{e^{(E + \mu _{0})/T_{0}} + 1}\right] 
 \int_{-p}^p dk J_m^+(q\rho).
 \label{eq:ppsi_aux}
\end{equation}
Taking into account the identity \eqref{eq:Jsum}, the sum over 
$m$ can be performed:
\begin{equation}
 \sum_{m = -\infty}^\infty J_m^+(q\rho) = 
 2\sum_{n = -\infty}^\infty J_n(q\rho) = 2,
 \label{eq:Jplus_sum}
\end{equation}
where $m = \pm \frac{1}{2}, \pm \frac{3}{2}, \dots$, while 
$n = 0, \pm 1, \pm 2, \dots$. After performing the sum over 
$m$ in \eqref{eq:ppsi_aux}, the integration variable can 
be changed from $E$ to $p$, giving:
\begin{equation}
 \braket{:\widehat{\overline{\Psi}}\widehat{\Psi}:}_{T_{0}} =
 \frac{M}{\pi^2} 
 \int_{0}^\infty \frac{dp\, p^2}{E}
 \left[\frac{1}{e^{(E - \mu _{0})/T_{0}} + 1} +
 \frac{1}{e^{(E + \mu _{0})/T_{0}} + 1}\right].
\end{equation}
The above expression coincides with 
that for $(E_F - 3P_F) / M$ \eqref{eq:RKT_gen_F}, obtained in RKT with $g_{\rm {F}} = 2$ (taking into account the 
fermion helicities) and $\Omega = 0$.
Thus, the FC has no corrections in the 
QFT setting compared to its RKT counterpart.

\subsection{Charge current} 
\label{sec:tevs0:CC}

The charge
conjugation property \eqref{eq:cc} can be used to show that:
\begin{equation}
 \mathfrak{J}^\mu(V_j, V_j) = \overline{U}_j^* \gamma^y \gamma^\mu \gamma^y U_j^* 
 = (\overline{U}_j \gamma^\mu U_j)^* = [\mathfrak{J}^\mu(U_j, U_{j'})]^*,
 \label{eq:JVV}
\end{equation}
where, as well as \eqref{eq:VV}, the 
properties $\gamma^y \gamma^\mu = 2\eta^{y\mu} - \gamma^\mu \gamma^y$ and $(\gamma^y)^*= -\gamma^y$ were used.
Thus, it is sufficient to compute $\mathfrak{J}^\mu(U_j, U_j)$.
Substituting $\mu = t$ and $\mu =i$ for the index $\mu$, we find:
\begin{equation}
 \mathfrak{J}^t(U_j, U_j) = \frac{1}{4\pi^2} \phi_j^\dagger \phi_j, \qquad 
 \mathfrak{J}^i(U_j, U_j) = \frac{1}{4\pi^2} \frac{2\lambda_j p_j}{E_j} 
 \phi_j^\dagger \sigma^i \phi_j. \qquad 
\end{equation}
It is convenient to work with components taken with respect to the tetrad 
introduced in \eqref{eq:frame_cyl}. The sigma matrices constructed 
with respect to this tetrad are:
\begin{equation}
 \sigma^{\hat{\rho}} = 
 \begin{pmatrix}
  0 & e^{-i\varphi} \\
  e^{i\varphi} & 0
 \end{pmatrix}, \qquad 
 \sigma^{\hat{\varphi}} = 
 \begin{pmatrix}
  0 & -i e^{-i\varphi} \\
  i e^{i\varphi} & 0
 \end{pmatrix}.
\end{equation}
The following relations can be established:
\begin{align}
 \phi_j^\dagger \phi_j =& \frac{1}{2} J_{m_j}^+(q_j \rho) + \frac{\lambda_j k_j}{p_j} J_{m_j}^-(q_j \rho), &
 \phi_j^\dagger \sigma^{\hat{\rho}} \phi_j =& 0, 
 \nonumber\\
 \phi_j^\dagger \sigma^{\hat{z}} \phi_j =& 
 \frac{1}{2} J_{m_j}^-(q_j \rho) + \frac{\lambda_j k_j}{p_j} J_{m_j}^+(q_j \rho), &
 \phi_j^\dagger \sigma^{\hat{\varphi}} \phi_j =& 
 \frac{\lambda_j q_j}{p_j} J_{m_j}^\times(q_j\rho),  
\end{align}
where the functions $J_m^\pm(q\rho)$ and 
$J_m^\times(q\rho)$ were introduced in \eqref{eq:Jstar_def}.

Noting that the density of states factors 
$[e^{(E \pm \mu _{0})/T_{0}} + 1]^{-1}$ 
are invariant under the transformation $k \rightarrow -k$,
$\lambda \rightarrow -\lambda$ and
$m \rightarrow -m$, it can be seen that the 
spatial components of $J^\mu$ vanish. This is 
because $J^-_m(q\rho)$ and $J_m^\times(q\rho)$ are 
odd with respect to $m \rightarrow -m$, while 
$\phi_j^\dagger \sigma^{\hat{z}} \phi_j$ is odd 
under the transformation 
$(k,m) \rightarrow (-k,-m)$. The time component of the 
CC can then be written as:
\begin{equation}
 \braket{:{\widehat {J}}^{\hat {t}}:}_{T_{0}} = \frac{1}{4\pi^2} \sum_{m = -\infty}^\infty 
 \int_M^\infty dE\, E 
 \left[\frac{1}{e^{(E - \mu _{0})/T_{0}} + 1} -
 \frac{1}{e^{(E + \mu _{0})/T_{0}} + 1}\right] 
 \int_{-p}^p dk J_m^+(q\rho).
\end{equation}
After performing the sum over $m$ using \eqref{eq:Jplus_sum},
an angle $\vartheta$ can be introduced
such that $k = p \cos\vartheta$ and $q = p \sin\vartheta$. 
The integration 
measure $E\, dE\, dk = q \, dq  \, dk$ is then changed to $p^2 \sin\vartheta\, d\vartheta \, dp$.
Since, after the sum over $m$ is performed, the integrand is independent 
of $\vartheta$, the integration with respect to this variable can be performed 
automatically, yielding $\int_0^\pi d\vartheta\, \sin\vartheta = 2$. 
Thus $\braket{:J^{\hat {t}}:}_{T_{0}}$ reduces to:
\begin{equation}
 \braket{:{\widehat{J}}^{\hat {t}}:}_{T_{0}} = \frac{1}{\pi^2} 
 \int_{0}^\infty dp\, p^2 
 \left[\frac{1}{e^{(E - \mu _{0})/T_{0}} + 1} -
 \frac{1}{e^{(E + \mu _{0})/T_{0}} + 1}\right].
\end{equation}
As was the case for the FC, the above expression
coincides with the fermion charge density $Q_{\rm {F}}$ \eqref{eq:RKT_QF}
obtained using RKT with $g_{\rm {F}} = 2$ and $\Omega = 0$, showing that there are no
quantum corrections.

\subsection{Stress-energy tensor} 
\label{sec:tevs0:SET}

In a manner similar to the one employed to derive \eqref{eq:JVV}, 
it can be shown that $\mathcal{T}_{\mu\nu}(V_j,V_j)$ can be
related to $\mathcal{T}_{\mu\nu}(U_j,U_j)$ via:
\begin{equation}
 \mathcal{T}_{\mu\nu}(V_j,V_j) = -\frac{i}{2}[
 \overline{U}_j \gamma_{(\mu} \partial_{\nu)} U_j - 
 \partial_{(\mu} \overline{U}_j \gamma_{\nu)} U_j]^* 
 = -[\mathcal{T}_{\mu\nu}(U_j, U_j)]^*,
\end{equation}
where the last $-$ sign comes from the complex conjugate of the 
imaginary unit $i$ prefactor. 
Using the properties \eqref{eq:Jprime} of the Bessel functions, we can derive the following relations:
\begin{align}
 \phi_j^\dagger \sigma^{\hat{\rho}} \partial_\rho \phi_j =& 
 \frac{i q_j^2 \lambda_j}{p_j} \left[J_{m_j}^+(q_j \rho) - 
 \frac{m_j}{q_j \rho} J_{m_j}^{\times}(q_j \rho)\right],
 \nonumber \\
 \phi_j^\dagger \sigma^{\hat{\varphi}} \partial_\varphi \phi_j =& 
 \frac{i m_j q_j \lambda_j}{p_j} J_{m_j}^\times(q_j \rho).
\end{align}
For stationary states, all off-diagonal tetrad components of the SET vanish.
However, when we consider rigidly-rotating states in the next section, the component $T_{\hat {t}{\hat {\varphi }}}$ will be nonzero.
We therefore write down the diagonal tetrad components and the $({\hat {t}}, {\hat {\varphi }})$ component which we will require later:
\begin{align}
 \mathcal{T}_{\hat{t}\hat{t}}(U_j, U_j) =& \frac{E_j}{8\pi^2} 
 \left[J_{m_j}^+(q_j \rho) + 
 \frac{2 \lambda_j k_j}{p_j} J_{m_j}^-(q_j \rho)\right],\nonumber\\
 \mathcal{T}_{\hat{t}\hat{\varphi}}(U_j, U_j) =& -\frac{1}{16\pi^2 \rho} 
 \left[\left(m_j - \frac{\lambda_j k_j}{p_j}\right) J_{m_j}^+(q_j \rho) -
 \left(\frac{1}{2} - 
 \frac{2 \lambda_j k_j m_j}{p_j}\right) J_{m_j}^-(q_j \rho)\right] \nonumber\\
 &-\frac{q_j}{16\pi^2} J_{m_j}^\times(q_j\rho),\nonumber\\
 \mathcal{T}_{\hat{\rho}\hat{\rho}}(U_j, U_j) =& \frac{q_j^2}{8\pi^2 E_j} 
 \left[J_{m_j}^+(q_j \rho) - 
 \frac{m_j}{q_j \rho} J_{m_j}^\times(q_j \rho)\right],\nonumber\\
 \mathcal{T}_{\hat{\varphi}\hat{\varphi}}(U_j, U_j) =& 
 \frac{q_j m_j}{8\pi^2 E_j \rho} J_{m_j}^\times(q_j \rho),\nonumber\\ 
 \mathcal{T}_{\hat{z}\hat{z}}(U_j, U_j) =& 
 \frac{k_j^2}{8\pi^2 E_j} J_{m_j}^+(q_j \rho) +
 \frac{\lambda_j k_j p_j}{4\pi^2 E_j} J_{m_j}^-(q_j \rho).
\end{align}
Using the summation formula:
\begin{equation}
 \sum_{m = -\infty}^\infty m J_m^\times(q\rho) = 
 \sum_{n = -\infty}^\infty (2n + 1) J_n(q\rho) J_{n + 1}(q\rho) 
 = 1,
\end{equation}
where, as before, $m = \pm \frac{1}{2}, \pm \frac{3}{2}, \dots$, while
$n= 0, \pm1 , \pm 2, \dots$, it can be shown that the t.e.v.~of the SET for nonrotating states
has the simple diagonal form
\begin{equation}
 \braket{:{\widehat {T}}_{\hat{\alpha}\hat{\sigma}}:}_{T_{0}} = 
 {\rm diag}(E_{F}, P_{F}, P_{F}, P_{F}),
\end{equation}
where 
$E_F$ and $P_F$ were obtained in \eqref{eq:RKT_gen_F} 
using the RKT formulation with $g_F = 2$. 
Therefore there are no quantum corrections to t.e.v.s for stationary states.

\section{Quantum rigidly-rotating thermal expectation values} \label{sec:tevs}

In the previous section, the construction of stationary thermal states 
was based on the nonrotating Minkowski vacuum, defined by setting the energy $E_{j}>0$ for particle modes. 
When the rotation is switched on, as discussed in section \ref{sec:therm},
we can define a rotating vacuum for fermions by instead setting the corotating energy ${\widetilde {E}}_{j}>0$ \eqref{eq:Etilde} to be positive for particle modes \cite{Iyer:1982ah}.
We therefore define the fermion field operator 
as follows:
\begin{multline}
 \widehat{\Psi} = \sum_{\lambda = \pm \frac{1}{2}} 
 \sum_{m = -\infty}^\infty 
 \int_{|E| > M} dE \, |E|\,
 \int_{-p}^p dk \, \Theta(\widetilde{E})\\\times
 \left[\hat{b}_{E,k,m}^\lambda U_{E,k,m}^\lambda(x) + 
 \hat{d}_{E,k,m}^\lambda{}^\dagger V_{E,k,m}^\lambda(x)\right] ,
 \label{eq:vacr}
\end{multline}
where the particle spinors $U_{E,k,m}^{\lambda }$ and anti-particle spinors $V_{E,k,m}^{\lambda }$ can be found in (\ref{eq:Ujfinal}, \ref{eq:Vjfinal}) respectively. 
The field operator \eqref{eq:vacr} should be compared with the corresponding definition \eqref{eq:vacs} for the stationary case.
In \eqref{eq:vacs} the integral over $E$ involves only positive energy $E>0$, whereas in \eqref{eq:vacr} we also take into account negative energy modes,
provided that the mass shell condition $|E| > M$ is satisfied.
Instead, the requirement that the co-rotating energy is positive, ${\widetilde {E}}>0$, is imposed by the presence of the Heaviside step function $\Theta ({\widetilde {E}})$.

With the decomposition (\ref{eq:vacr}) of the fermion field operator, we can proceed to construct t.e.v.s  using the method employed in section~\ref{sec:tevs0} in the stationary case.
The mode expansion \eqref{eq:vacr} is inserted into the FC, CC and SET operators \eqref{eq:schematic}, to obtain mode sums involving the particle and anti-particle creation and annihilation operators. 
The t.e.v.s of the particle number operators are then given by \eqref{eq:bblocks}, where the temperature on the axis of rotation is fixed to be $T_{0}$. 
The density of states factor in \eqref{eq:bblocks} now has a dependence on the angular momentum quantum number $m_{j}$ as well as the energy $E_{j}$.
In this section we study the t.e.v.s of the 
FC, CC and AC for a rigidly-rotating thermal state. 
We consider the SET separately in section~\ref{sec:SET}.

\subsection{Fermion condensate}
\label{sec:tevs:FC}

Starting from \eqref{eq:vacr}, the following expression 
is obtained for the t.e.v. of the FC:
\begin{multline}
 \braket{:\widehat{\overline{\Psi}}\widehat{\Psi}:}_{T_{0}} =
 \sum_{\lambda = \pm \frac{1}{2}} 
 \sum_{m = -\infty}^\infty 
 \int_{|E| > M} dE \, |E|\,
 \int_{-p}^p dk \, \Theta(\widetilde{E})
 \\\times
 \left[\frac{\overline{U}^\lambda_{E,k,m} U^\lambda_{E,k,m}}
 {e^{(\widetilde{E} - \mu_0)/T_0} + 1} - 
 \frac{\overline{V}^\lambda_{E,k,m} V^\lambda_{E,k,m}}
 {e^{(\widetilde{E} + \mu_0)/T_0} + 1}\right].
\end{multline}
Using (\ref{eq:VV}, \ref{eq:UU}), the sum over $\lambda$
can be performed, yielding:
\begin{multline}
 \braket{:\widehat{\overline{\Psi}}\widehat{\Psi}:}_{T_{0}} =
 \frac{M}{4\pi^2} \sum_{m = -\infty}^\infty 
 \int_{|E| > M} dE \, {\rm sgn}(E)
 \int_{-p}^p dk \, \Theta(\widetilde{E}) \, J_m^+(q\rho) 
 \\\times
 \left[\frac{1}{e^{(\widetilde{E} - \mu_0)/T_0} + 1} + 
 \frac{1}{e^{(\widetilde{E} + \mu_0)/T_0} + 1}\right],
\end{multline}
where ${\rm sgn}(E) = |E| / E$ is the sign of the energy 
of the mode.
To simplify the integration above, the integral over $E$ can 
be split into its positive ($E > M$) and negative ($E < -M$) 
domains. On the negative branch, the simultaneous sign flip 
$(E, m) \rightarrow (-E,-m)$ can be performed, under which 
$\widetilde{E} \rightarrow -\widetilde{E}$. Noting that
$J_{-m}^+(q\rho) = J_m^+(q\rho)$, the following expression is
obtained:
\begin{multline}
 \braket{:\widehat{\overline{\Psi}}\widehat{\Psi}:}_{T_{0}} =
 \frac{M}{4\pi^2} \sum_{m = -\infty}^\infty 
 \int_{M}^\infty dE \int_{-p}^p dk \, 
 J_m^+(q\rho) \, {\rm sgn}(\widetilde{E}) \\ \times
 \left[\frac{1}{e^{(|\widetilde{E}| - \mu_0)/T_0} + 1} + 
 \frac{1}{e^{(|\widetilde{E}| + \mu_0)/T_0} + 1}\right].
\end{multline}
In order to study the massless limit of $M^{-1}\braket{:\widehat{\overline{\Psi}}\widehat{\Psi}:}_{T_{0}}$, we now attempt to 
simplify the integrand,  
by replacing ${\rm sgn}(\widetilde{E}) = 1$ 
and $|\widetilde{E}| = \widetilde{E}$. 
To this end, consider the quantity ${\mathfrak {F}}_{1}$
\begin{equation}
 {\mathfrak {F}}_{1} = \sum_{m = -\infty}^\infty \int_M^\infty dE 
 \left[\frac{{\rm sgn}(\widetilde{E})}
 {e^{(|\widetilde{E}| - \mu _{0})/T_{0}} + 1} +
 \frac{{\rm sgn}(\widetilde{E})}
 {e^{(|\widetilde{E}| + \mu _{0})/T_{0}} + 1}\right] 
 {\mathfrak {f}}(m,E),
\end{equation}
where ${\mathfrak {f}}(m,E)$ is a function depending on $m$ and $E$,
We now write ${\mathfrak {F}}_{1}$ as a sum of a term $\left\{ {{\mathfrak {F}}}_{1}\right\} _{\rm {simp}}$ where 
$|\widetilde{E}|$ is replaced by $\widetilde{E}$ (that is, the 
modulus is removed) and ${\rm sgn}(\widetilde{E})$ is set equal to one, and a remainder $\Delta {\mathfrak {F}}_{1}$:
\begin{equation}
 {\mathfrak {F}}_{1} = \left\{ {{\mathfrak {F}}}_{1}\right\} _{\rm {simp}} + \Delta {\mathfrak {F}}_{1},
 \end{equation}
where
\begin{align}
  \left\{ {{\mathfrak {F}}}_{1}\right\} _{\rm {simp}} =& \sum_{m = -\infty}^\infty \int_M^\infty dE 
 \left[\frac{1}{e^{(\widetilde{E} - \mu _{0})T_{0}} + 1} +
 \frac{1}{e^{(\widetilde{E} + \mu _{0})/T_{0}} + 1}\right]
  {\mathfrak {f}}(m,E), \nonumber\\
 \Delta {\mathfrak {F}}_{1} =& -\sum_{m = m_M}^\infty \int_M^{\Omega m} dE 
 \left[\frac{1}{e^{(-\widetilde{E} - \mu _{0})/T_{0}} + 1} +
 \frac{1}{e^{(-\widetilde{E} + \mu _{0})/T_{0}} + 1}
 \right.\nonumber\\ 
 & \left. \hspace{70pt}+
 \frac{1}{e^{(\widetilde{E} - \mu _{0})/T_{0}} + 1} +
 \frac{1}{e^{(\widetilde{E} + \mu _{0})/T_{0}} + 1}
 \right] {\mathfrak {f}}(m,E)\nonumber\\
 =& -\sum_{m = m_M}^\infty \int_M^{\Omega m} dE \, 2 {\mathfrak {f}}(m,E), 
\end{align}
where $m_M$ is the minimum value of $m$ for which $\Omega m > M$.
The last line follows from the identity
$(e^x + 1)^{-1} + (e^{-x} + 1)^{-1} = 1$. 
The last equality above shows that $\Delta {\mathfrak {F}}_{1}$ does not depend on 
$T_{0}$ or $\mu _{0}$ unless ${\mathfrak {f}}(m,E)$ explicitly 
depends on these parameters (which it does not for the FC). 
The dependence of $\Delta {\mathfrak {F}}_{1}$ on 
$\Omega$ is due to the definition of the rotating 
vacuum, where $\Omega$ appears explicitly when restricting the 
energy spectrum to positive co-rotating energies.
We thus find 
\begin{multline}
 M^{-1} \left\{ \braket{:\widehat{\overline{\Psi}}\widehat{\Psi}:}_{T_{0}}\right\} _{\rm {simp}} = 
 \frac{1}{2\pi^2} 
 \sum_{m = -\infty}^\infty 
 \int_M^\infty dE
 \left[ 
 \frac{1}{e^{(\widetilde{E} - \mu_0) / T_0} + 1} +
 \frac{1}{e^{(\widetilde{E} + \mu_0) / T_0} + 1}\right]
 \\\times 
 \int_{0}^p dk\, J_{m}^+(q\rho).
\end{multline}
In the massless limit, the following exact result can be obtained (see \cite{Ambrus:2019ayb,Ambrus:2019khr} for further details of the techniques used to  perform the integration):
\begin{equation}
 \left.M^{-1} \left\{ \braket{:\widehat{\overline{\Psi}}\widehat{\Psi}:}_{T_{0}}\right\} _{\rm {simp}}  \right\rfloor_{M = 0} = 
 \frac{T^2}{6} + \frac{\mu^2}{2\pi^2} + 
 \frac{3\bm{\omega}^2 + 2 \bm{a}^2}{24\pi^2},
 \label{eq:ppsi_bar}
\end{equation}
where $\bm{\omega}^2 = \Omega^2 \Gamma^2$ and $\bm{a}^2 = \rho^2 \Omega^2 \Gamma^4$ are the squares of the spatial parts of the kinematic vorticity and acceleration introduced in \eqref{eq:kinematic}, while
$\Gamma $ is the Lorentz factor \eqref{eq:RKT_sol}.
The last term is independent of 
$\mu$ and $T$
and hence 
represents the contribution due to the difference between the 
rotating and stationary vacua. Subtracting this contribution 
gives 
\begin{equation}
 \left.M^{-1} \braket{:\widehat{\overline{\Psi}}\widehat{\Psi}:}_{T_0} \right\rfloor_{M = 0} = 
\frac{T^2}{6} + \frac{\mu^2}{2\pi^2},
 \label{eq:ppsi}
\end{equation}
which agrees with the  RKT result \eqref{eq:RKT_M0tr} with $g_{\rm {F}}=2$, diverging as $\Gamma \rightarrow \infty $ and the SLS is approached.

\subsection{Charge current} 
\label{sec:tevs:CC}

Since the density of states factor in \eqref{eq:bblocks} now has a dependence on the angular momentum 
quantum number $m$ as well as the energy $E$, the ${\hat {\varphi}}$ component of 
the CC no longer vanishes when the state is rigidly-rotating.
The nonzero components of the t.e.v.~of the CC take the form:
\begin{multline}
 \begin{pmatrix}
 \braket{:\!{\widehat{J}}^{\hat{t}}\!:}_{T_{0}} \\
 \braket{:\!{\widehat{J}}^{\hat{\varphi}}\!:}_{T_{0}} 
 \end{pmatrix}
 = \frac{1}{4\pi^2} \sum_{m = -\infty}^\infty 
 \int_M^\infty dE
 \left[\frac{1}{e^{(|\widetilde{E}| - \mu _{0})/T_{0}} + 1} -
 \frac{1}{e^{(|\widetilde{E}| + \mu _{0})/T_{0}} + 1}\right] \\\times
 \int_{-p}^p dk 
 \begin{pmatrix}
  E\, J_m^+(q\rho) \\
  q\, J_m^\times(q\rho)
 \end{pmatrix} .
 \label{eq:CC_aux}
\end{multline}
To compute the above integrals in the massless limit, we follow the method employed for the FC and define a quantity
\begin{equation}
 {\mathfrak {F}}_{2} = \sum_{m = -\infty}^\infty \int_M^\infty dE 
 \left[\frac{1}{e^{(|\widetilde{E}| - \mu _{0})/T_{0}} + 1} -
 \frac{1}{e^{(|\widetilde{E}| + \mu _{0})/T_{0}} + 1}\right]
  {\mathfrak {f}}(m,E).
\end{equation}
Writing ${\mathfrak {F}}_{2}$ as a sum of  a term  $\left\{ {{\mathfrak {F}}}_{2}\right\} _{\rm {simp}}$ where 
$|\widetilde{E}|$ is replaced by $\widetilde{E}$ and a remainder $\Delta {\mathfrak {F}}_{2}$
\begin{equation}
 {\mathfrak {F}}_{2} = \left\{ {{\mathfrak {F}}}_{2}\right\} _{\rm {simp}} + \Delta {\mathfrak {F}}_{2},
 \end{equation}
we find
\begin{align}
 \left\{ {{\mathfrak {F}}}_{2}\right\} _{\rm {simp}} =& \sum_{m = -\infty}^\infty \int_M^\infty dE 
 \left[\frac{1}{e^{(\widetilde{E} - \mu _{0})T_{0}} + 1} -
 \frac{1}{e^{(\widetilde{E} + \mu _{0})/T_{0}} + 1}\right]
  {\mathfrak {f}}(m,E), \nonumber\\
 \Delta {\mathfrak {F}}_{2} =& \sum_{m = m_M}^\infty \int_M^{\Omega m} dE 
 \left[\frac{1}{e^{(-\widetilde{E} - \mu _{0})/T_{0}} + 1} -
 \frac{1}{e^{(-\widetilde{E} + \mu _{0})/T_{0}} + 1}\right. 
 \nonumber\\
 &\left.\hspace{70pt} - 
 \frac{1}{e^{(\widetilde{E} - \mu _{0})/T_{0}} + 1} +
 \frac{1}{e^{(\widetilde{E} + \mu _{0})/T_{0}} + 1}\right]
  {\mathfrak{f}}(m,E).
\end{align}
The term inside the square brackets in 
$\Delta {\mathfrak {F}}_{2}$ is 
identically zero. 
Thus, it can be concluded that ${\mathfrak {F}}_{2} =  \left\{ {{\mathfrak {F}}}_{2}\right\} _{\rm {simp}}$  
for any function ${\mathfrak {f}}(m,E)$, which simplifies the integration.

The following expressions 
are then obtained for massless fermions \cite{Ambrus:2019ayb,Ambrus:2019khr}:
\begin{align}
 \braket{:{\widehat{J}}^{\hat{t}}:}_{T_{0}} =& 
 \frac{\mu _{0}\Gamma ^{4}}{3} \left(T_{0}^{2} + 
 \frac{\mu _{0}^{2}}{\pi^2}\right)
 + 
 \frac{\mu _{0} \Omega^2 \Gamma^4}{4\pi^2} \left(
 \frac{4}{3}\Gamma^2 - \frac{1}{3}\right)
 = \Gamma \left[Q_F + 
 \frac{\mu}{12\pi^2}(3\bm{\omega}^2 + \bm{a}^2)\right],\nonumber\\
 \braket{:{\widehat{J}}^{\hat{\varphi}}:}_{T_{0}}=& 
 \rho \Omega \Gamma \left[Q_F + 
 \frac{\mu}{12\pi^2}(\bm{\omega}^2 + 3 \bm{a}^2)\right]. \label{eq:CCr}
\end{align}
As expected, the $\varphi$-component vanishes when $\Omega =0$ and the state is nonrotating.
The first terms appearing on the right-hand-side correspond to the RKT results for $g_{\rm {F}} = 2$ \eqref{eq:RKT_M0}.
The second terms are the quantum corrections, and are proportional to $\Omega ^{2}$, vanishing when the rotation is zero.
The quantum corrections do not depend on the temperature $T$, only on the chemical potential, 
local vorticity and local acceleration.
The quantum corrections are therefore present even in the zero-temperature limit. 
The decomposition of the CC with respect to the kinematic tetrad in \eqref{eq:kinematic} will be discussed in section \ref{sec:beta}.

The t.e.v.~of the CC vanishes identically when the chemical potential on the axis $\mu _{0}$ is zero.
This is to be expected since, with vanishing chemical potential, a rigidly-rotating thermal state will contain equal numbers of particles and anti-particles.
When $\mu _{0}$ is nonzero, the current diverges as $\Gamma \rightarrow \infty $ and the SLS is approached. 
For both components of the CC, the quantum corrections diverge more rapidly than the RKT contributions as $\rho \rightarrow \Omega ^{-1}$. 
Therefore,  close to the SLS, the CC is completely dominated by quantum effects and the RKT contributions are subleading.

\subsection{Axial current}
\label{sec:tevs:J5}

The classical AC $J_{5}^{\mu }$ is defined by 
\begin{equation}
 J^\mu_5 = \overline{\psi} \gamma^\mu \gamma_5 \psi ,
 \label{eq:J5_def}
\end{equation}
where we have introduced the chirality matrix
\begin{equation}
 \gamma_5 = i \gamma^t \gamma^x \gamma^y \gamma^z =
 \begin{pmatrix}
  0 & 1 \\ 1 & 0
 \end{pmatrix} .
\end{equation}
Using the 
Dirac equation \eqref{eq:dirac}, and taking into account 
that $\gamma_5$ anti-commutes with all of the other 
$\gamma$ matrices, $\{\gamma_5, \gamma^\mu\} = 0$, we find
$\partial_\mu J^\mu_5 = 2 i M \overline{\psi} \gamma_5 \psi$,
and hence $J^\mu_5$ is conserved for massless particles. 
Nonvanishing values of $J^\mu_5$ can 
be induced through the chiral vortical effect 
(for a review, see \cite{Kharzeev:2016}). 
The expectation 
values of $J^\mu_5$ computed for massless fermions using 
a perturbative approach were recently reported in \cite{Buzzegoli:2018}.
Here we consider the t.e.v.~of $J^\mu_5$ using
QFT techniques.

Using the mode expansion \eqref{eq:vacr}, the t.e.v.~of \eqref{eq:J5_def} takes the form:
\begin{align}
 \braket{:\widehat{J}^\mu_5:}_{T_0} =& \sum_j \left\{ 
 \frac{\mathfrak{J}^\mu_5(U_j, U_j)}
 {\exp[(\widetilde{E}_j - \mu_0) / T_0] + 1} -
 \frac{\mathfrak{J}^\mu_5(V_j, V_j)}
 {\exp[(\widetilde{E}_j + \mu_0) / T_0] + 1}\right\},
 \label{eq:J5_tev_aux}
 \end{align}
 where $\mathfrak{J}^\mu_5(\psi,\chi) = 
 \overline{\psi} \gamma^\mu \gamma_5 \chi$.
Following the same reasoning applied to obtain \eqref{eq:JVV},
it is not difficult to show that 
$\mathfrak{J}^\mu_5(V_j, V_j) ={\color{blue} -} [\mathfrak{J}^\mu_5(U_j, U_j)]^*$,
while
\begin{align}
 \mathfrak{J}^t_5(U_j, U_j) =& \frac{p_j}{8\pi^2 E_j}
 \left[2\lambda_j J_{m_j}^+(q_j \rho) + 
 \frac{k_j}{p_j} J_{m_j}^-(q_j\rho)\right],\nonumber\\
 \mathfrak{J}^\varphi_5(U_j, U_j) =& 
 \frac{\lambda_j q_j}{4\pi^2 p_j} J_{m_j}^\times(q_j\rho),
 \nonumber\\
 \mathfrak{J}^z_5(U_j, U_j) =& \frac{1}{8\pi^2} \left[ 
 J_{m_j}^-(q_j \rho) + \frac{2\lambda_j k_j}{p_j} 
 J_{m_j}^+(q_j \rho)\right].
\end{align}
When considering the sum over $j$ in \eqref{eq:J5_tev_aux},
the terms which are odd with respect to $\lambda$ and 
$k$ vanish. Thus, the only non-vanishing component of the 
t.e.v.~of the AC is
\begin{multline}
 \braket{:\widehat{J}^z_5:}_{T_0} = \frac{1}{4\pi^2} 
 \sum_{m = -\infty}^\infty 
 \int_M^\infty dE\, E \int_{-p}^p dk\,
 J_{m}^-(q\rho) {\rm sgn}(\widetilde{E})\\
 \times \left\{ 
 \frac{1}
 {\exp[(|\widetilde{E}| - \mu_0) / T_0] + 1} +
 \frac{1}
 {\exp[(|\widetilde{E}| + \mu_0) / T_0] + 1}\right\}.
\end{multline}

As in the cases of the FC and CC, 
the t.e.v.~of the axial current 
can be computed exactly in the massless limit.
We simplify as discussed in section~\ref{sec:tevs:FC},
replacing ${\rm sgn}(\widetilde{E}) = 1$ and 
$|\widetilde{E}| = \widetilde{E}$, to find:
\begin{multline}
 \left\{ \braket{:\widehat{J}^z_5:}_{T_0} \right\} _{\rm {simp}} = 
 \frac{1}{2\pi^2} 
 \sum_{m = -\infty}^\infty 
 \int_M^\infty dE\, E 
 \left[ 
 \frac{1}{e^{(\widetilde{E} - \mu_0) / T_0} + 1} +
 \frac{1}{e^{(\widetilde{E} + \mu_0) / T_0} + 1}\right]
 \\\times \int_{0}^p dk\, J_{m}^-(q\rho).
\end{multline}
In the massless limit, the following exact result can be obtained \cite{Ambrus:2019ayb,Ambrus:2019khr}:
\begin{align}
 \left.\left\{ \braket{:\widehat{J}^z_5:}_{T_0} \right\} _{\rm {simp}}
 \right\rfloor_{M = 0} =& 
 \frac{\Omega T_0^2 \Gamma^4}{6}
 \left(1 + \frac{3\mu_0^2}{\pi^2 T_0^2}\right) + 
 \frac{\Omega^3 \Gamma^4}{24\pi^2} (4\Gamma^2 - 3)\nonumber\\
 =& \omega^{\hat{z}} \left(
 \frac{T^2}{6} + \frac{\mu^2}{2\pi^2} + 
 \frac{\bm{\omega}^2 + 3\bm{a}^2}{24\pi^2}\right),
 \label{eq:J5z_bar}
\end{align}
where $\omega^{\hat{\alpha}}$ is the kinematic vorticity introduced in \eqref{eq:kinematic}.
The last term is independent of $\mu_0$ and $T_0$ and hence 
represents the contribution due to the difference between the 
rotating and stationary vacua \cite{Ambrus:2014uqa}. 
Eliminating this term allows 
the t.e.v. of the AC to be obtained as:
\begin{equation}
 \left.\braket{:\widehat{J}^{\hat{\alpha}}_5:}_{T_0}\right\rfloor_{M = 0} = \sigma^\omega_A \omega^{\hat{\alpha}}, \qquad 
 \sigma^\omega_A = \frac{T^2}{6} + \frac{\mu^2}{2\pi^2},
 \label{eq:J5z}
\end{equation}
where $\sigma^\omega_A$ is the axial vortical conductivity, which allows an axial charge flow to develop along the kinematic vorticity vector.
As expected, the AC \eqref{eq:J5z} vanishes in the stationary case, but, unlike the CC, it is nonzero even when the chemical potential vanishes \cite{Ambrus:2014uqa}.

The AC vanishes in classical RKT.
Restoring the reduced Planck's constant, the AC \eqref{eq:J5z} is proportional to $\hbar \Omega $ and is therefore larger than the quantum corrections to the CC, which are ${\mathcal {O}}(\hbar ^{2}\Omega ^{2})$.

The AC has been studied previously by a number of authors \cite{Kharzeev:2016,Prokhorov:2018,Vilenkin:1979}.
Up to possible overall factors due to differences in definitions, \eqref{eq:J5z} agrees with the corresponding quantity in \cite{Kharzeev:2016} only on the rotation axis, where $\Gamma =1$.
The axial current in \cite{Prokhorov:2018} (derived using the ansatz for the Wigner function proposed in \cite{Becattini:2013})
matches \eqref{eq:J5z_bar} only on the axis of rotation, but no distinction is made in \cite{Prokhorov:2018} between the stationary and rotating vacua.
Constructed using a QFT approach and considering the stationary Minkowski vacuum, the AC in \cite{Vilenkin:1979} agrees with \eqref{eq:J5z_bar}, again only on the axis of rotation. 
Finally, the result obtained in \cite{Prokhorov:2018bql} using perturbative QFT agrees fully with \eqref{eq:J5z_bar}.

\section{Hydrodynamic analysis of the quantum stress-energy tensor}
\label{sec:SET}

In this section we consider in detail the t.e.v.~of the SET for rigidly-rotating states.
Following the approach of the previous section, we first derive the components of this t.e.v.~with respect to the orthonormal tetrad \eqref{eq:frame_cyl}.
For comparison with the RKT results from section~\ref{sec:RKT}, we then consider quantities defined with respect to the $\beta $-frame (or thermometer frame). 

\subsection{Stress-energy tensor expectation values} 
\label{sec:tevs:SET}

The t.e.v.~of the SET can be written compactly as:
\begin{multline}
 \braket{:\widehat{T}_{\hat{\alpha}\hat{\sigma}}:}_{T_{0}} =
 \frac{1}{4\pi^2} \sum_{m = -\infty}^\infty 
 \int_{M}^\infty dE\, E \int_{-p}^p dk\, {\mathfrak {T}}_{\hat{\alpha}\hat{\sigma}} \, {\rm sgn}(\widetilde{E}) \\
 \times 
 \left[\frac{1}
 {e^{(|\widetilde{E}| - \mu _{0})/T_{0}} + 1} +
 \frac{1}
 {e^{(|\widetilde{E}| + \mu _{0})/T_{0}} + 1}\right],
 \label{eq:SET_frame}
\end{multline}
where the tensor ${\mathfrak {T}}_{\hat{\alpha}\hat{\sigma}}$ has the following non-vanishing components:
\begin{gather}
 {\mathfrak {T}}_{\hat{t}\hat{t}} = E\, J_m^+(q\rho), \qquad 
 {\mathfrak {T}}_{\hat{t}\hat{\varphi}} = -\frac{1}{2\rho} 
 \left[m J_m^+(q \rho) - \frac{1}{2} J_m^-(q\rho)\right] - \frac{q}{2} J_m^\times(q\rho),\nonumber\\
 {\mathfrak {T}}_{\hat{\rho}\hat{\rho}} = \frac{q^2}{E} \left[ J_m^+(q\rho) -  \frac{m}{q\rho} J_{m}^\times (q\rho)\right], \,\,\,
 {\mathfrak {T}}_{\hat{\varphi}\hat{\varphi}} = \frac{m q}{\rho E} J_m^\times(q\rho), \,\,\,
 {\mathfrak {T}}_{\hat{z}\hat{z}} = \frac{k^2}{E} J_m^+(q\rho).
 \nonumber \label{eq:SET_aux}\\
\end{gather}
As with the t.e.v.s considered in section~\ref{sec:tevs}, we can obtain closed-form expressions in the massless limit. 
We first simplify using the approach of section~\ref{sec:tevs:FC}, and then integrate using a procedure whose details
can be found in \cite{Ambrus:2019ayb,Ambrus:2019khr}. 
The results are:
\begin{align}
 \braket{:\widehat{T}_{\hat{t}\hat{t}}:}_{T_{0}} =& 
 P_{\rm F}(4\Gamma^2 - 1)
+ \frac{\Omega^2\Gamma^2}{8} \left(
 T^{2} + \frac{3\mu^2}{\pi^2} \right)
 \left(\frac{8}{3} \Gamma^4 - \frac{16}{9} \Gamma^2 + \frac{1}{9}\right),\nonumber\\
 \braket{:\widehat{T}_{\hat{t}\hat{\varphi}}:}_{T_{0}} =& -\rho \Omega \Gamma^2\Bigg[
 4 P_{\rm F}^2  + \frac{2 \Omega^2 \Gamma^2}{9} \left(
 T^{2} + \frac{3\mu^{2}}{\pi^2} \right)
 \left(\frac{3}{2} \Gamma^2 - \frac{1}{2}\right) \Bigg],\nonumber
\\
 \braket{:\widehat{T}_{\hat{\rho}\hat{\rho}}:}_{T_{0}} =& P_{\rm F} + \frac{\Omega^2 \Gamma^2}{24} \left(
 T^{2} + \frac{3\mu^{2}}{\pi^2} \right)
 \left(\frac{4}{3} \Gamma^2 - \frac{1}{3}\right),\nonumber\\
 \braket{:\widehat{T}_{\hat{\varphi}\hat{\varphi}}:}_{T_{0}} =& P_{\rm F}(4\Gamma^2 - 3)
 + \frac{\Omega^2 \Gamma^2}{24} \left(
 T^{2} + \frac{3\mu^{2}}{\pi^2} \right)
 \left(8\Gamma^4 - 8 \Gamma^2 + 1\right),\label{eq:SETr}
\end{align}
while $\braket{:\widehat{T}_{\hat{z}\hat{z}}:}_{T_{0}} = 
\braket{:\widehat{T}_{\hat{\rho}\hat{\rho}}:}_{T_{0}}$ (this relation holds also in the case of massive field quanta \cite{Ambrus:2019ayb,Ambrus:2014uqa}).
The first term in each component of the SET is the contribution from RKT (see section~\ref{sec:RKT}), while the second term is the quantum correction.
As for the CC (see section~\ref{sec:tevs:CC}), the quantum corrections are all proportional to $\Omega ^{2}$ and, as expected from section~\ref{sec:tevs0}, vanish in the stationary case.
Unlike the CC, the quantum corrections are now temperature-dependent.
All components of the t.e.v.~of the SET diverge on the SLS, and, once again, the quantum corrections diverge more quickly as $\Gamma \rightarrow \infty $.

\subsection{Thermometer frame}
\label{sec:beta}

Further insight into the effect of quantum corrections can be gleaned from a hydrodynamic analysis of the SET.
In relativistic fluid dynamics, the equivalence between mass and energy transfer 
makes the macroscopic four-velocity $u^{\mu }$ an ambiguous concept. 
A frame is defined by making a choice for 
the definition of $u^\mu$. 
Here we work in the $\beta$-frame, also termed the natural frame \cite{Van:2012}, or thermometer frame \cite{Van:2013}  (see also \cite{Becattini:2015} for an analysis of the properties of this frame).
In the $\beta$-frame, the macroscopic four-velocity $u^{\mu} $ is proportional to the temperature four-vector $\beta ^{\mu }$, that is, $u^{\mu } = T\beta ^{\mu }$, where $T$ is the local temperature.
For rigidly-rotating states, the macroscopic four-velocity is then given by \eqref{eq:4velocity}.

With this definition of $u^{\mu}$, we decompose the CC and SET as follows
\cite{Bouras:2010}:
\begin{equation}
 J^\mu = Q_{\beta } u^\mu + \mathcal{J}^\mu_\beta, \qquad 
 T^{\mu\nu} = E_{\beta }\, u^\mu u^\nu - (P_{\beta } + \varpi)\Delta^{\mu\nu} + 
 \Pi^{\mu\nu} + u^\mu W^\nu + u^\nu W^\mu,
 \label{eq:SET_dec}
\end{equation}
where $Q_{\beta }$, $E_{\beta }$ and $P_{\beta }$ are the usual equilibrium quantities,
$\mathcal{J}^\mu$ and $W^\mu$ represent the charge and heat flux 
in the local rest frame, $\varpi$ is the dynamic pressure
and $\Pi^{\mu\nu}$ is the pressure deviator. The tensor 
$\Delta^{\mu\nu} = g ^{\mu\nu} - u^\mu u^\nu$ is a projector on the 
hypersurface orthogonal to $u^\mu$. The nonequilibrium quantities 
$\mathcal{J}^\mu$, $\Pi^{\mu\nu}$ and $W^\mu$ are also orthogonal 
to $u^\mu$, by construction.
The isotropic pressure $P_{\beta } + \varpi$ 
is given as the sum of the hydrostatic pressure $P_{\beta }$,  computed using 
the equation of state of the fluid, and of the dynamic pressure $\varpi$,
which in general depends on the divergence of the velocity. 
In the case of massless (or ultrarelativistic) particles, the SET is traceless, since the massless Dirac field is conformally coupled and the conformal trace anomaly vanishes on flat space-time \cite{Duff:1993wm}.
From \eqref{eq:SET_dec}, the SET trace is $T^{\mu}{}_\mu = E_{\beta } - 3(P_{\beta } + \varpi$, and therefore 
$\varpi$ vanishes for massless particles since $E_{\beta }= 3P_\beta $. Moreover, since the velocity field is divergenceless ($\nabla_\mu u^\mu = 0$), it is reasonable to assume that $\varpi = 0$ also when $M > 0$. However, below we keep this term for clarity.

For both massive and massless particles, the macroscopic quantities can be extracted from the components of $J^\mu$ and $T^{\mu\nu}$ as follows \cite{Cercignani:2002}:
\begin{gather}
 Q_{\beta } = u_\mu J^\mu, \qquad E_{\beta } = u_\mu u_\nu T^{\mu\nu}, \qquad 
 P_{\beta }+\varpi = -\frac{1}{3} \Delta_{\mu\nu} T^{\mu\nu}, \nonumber\\
 \mathcal{J}^\mu = \Delta^{\mu\nu} J_\nu, \qquad 
 W^\mu = \Delta^{\mu\nu} u^\lambda T_{\nu\lambda}, \qquad 
 \Pi^{\mu\nu} = T^{\braket{\mu\nu}},
 \label{eq:SET_dec_inv}
\end{gather}
where the notation $A^{\braket{\mu\nu}}$ for a general two-index tensor denotes
\begin{equation}
 A^{\braket{\mu\nu}} = \left[\frac{1}{2} \left(
 \Delta^{\mu\lambda} \Delta^{\nu\sigma} + 
 \Delta^{\nu\lambda} \Delta^{\mu\sigma}\right)-
 \frac{1}{3} \Delta^{\mu\nu} \Delta^{\lambda\sigma}\right] 
 A_{\lambda\sigma}.
 \label{eq:SET_shear}
\end{equation}
Since in general, $\mathcal{J}^{\hat{\rho}} = \mathcal{J}^{\hat{z}} = 0$ and $\mathcal{J}^{\hat{\alpha}} u_{\hat{\alpha}} = 0$, it can be seen that $\mathcal{J}^{\hat{\alpha}}$ points along the circular vector $\tau^{\hat{\alpha}}$, introduced in \eqref{eq:kinematic}:
\begin{equation}
 \mathcal{J}^{\hat{\alpha}} = \sigma_V^\tau \tau^{\hat{\alpha}}, \qquad 
 \sigma_V^\tau = \frac{\rho \Omega J^{\hat{t}} - J^{\hat{\varphi}}}{\rho \Omega^3 \Gamma^3},
\end{equation}
where $\sigma_V^\tau$ is the circular vector (electric) charge conductivity.
Similarly, the structure of $T^{\mu\nu}$ indicates 
that $W^{\hat{\rho}} = W^{\hat{z}} = 0$, while 
the orthogonality between $W^{\hat{\alpha}}$ and 
$u^{\hat{\alpha}}$ allows $W^{\hat{\alpha}}$ to 
be written as:
\begin{equation}
 W^{\hat{\alpha}} = \sigma_\varepsilon^\tau \tau^{\hat{\alpha}}, \qquad 
 \sigma_\varepsilon^\tau = \frac{1}{\Omega^2 \Gamma^2} \left(T_{\hat{t}\hat{t}} + T_{\hat{\varphi}\hat{\varphi}} + \frac{1 + \rho^2 \Omega^2}{\rho\Omega} T_{\hat{t}\hat{\varphi}}\right),
\end{equation}
where $\sigma^\tau_\varepsilon$ is the circular heat conductivity.
Finally, noting that $\Pi^{\hat{\alpha}\hat{\sigma}}$ 
is symmetric, traceless and  orthogonal to $u^{\hat{\alpha}}$ with respect to both 
indices,
as well as the property $T^{\hat{\rho}\hat{\rho}} = T^{\hat{z}\hat{z}}$, only one degree of freedom is required to characterise $\Pi^{\hat{\alpha}\hat{\sigma}}$, introduced as $\Pi_\beta$ below \cite{Ambrus:2019ayb,Ambrus:2019khr}:
\begin{align}
 \Pi^{\hat{\alpha}\hat{\sigma}} =& 
 \Pi_\beta\left(\tau^{\hat{\alpha}} \tau^{\hat{\sigma}} - 
 \frac{\bm{\omega}^2}{2} a^{\hat{\alpha}} a^{\hat{\sigma}} - \frac{\bm{a}^2}{2} \omega^{\hat{\alpha}} \omega^{\hat{\sigma}}\right) = 
 \rho^2 \Omega^6 \Gamma^8 \Pi_\beta  
 \begin{pmatrix}
  \rho^2 \Omega^2 \Gamma^2 & 0 & \rho \Omega \Gamma^2 & 0\\
  0 & -\frac{1}{2} & 0 & 0\\
  \rho \Omega \Gamma^2 & 0 & \Gamma^2 & 0 \\
  0 & 0 & 0 & -\frac{1}{2}
 \end{pmatrix},\nonumber\\ 
 \Pi_\beta =& \frac{2(P_\beta + \varpi - T_{\hat{z}\hat{z}})}{\rho^2 \Omega^6 \Gamma^8}.
 \label{eq:Pi}
\end{align}

In the case of massless fermions, substituting the SET components \eqref{eq:SETr} 
into the contractions \eqref{eq:SET_dec_inv} yields the following 
closed form results:
\begin{align}
 Q_\beta =& Q_{\rm F} + \Delta Q, &
 Q_{\rm F} =& \frac{\mu}{3}\left(T^2 + 
 \frac{\mu^2}{\pi^2}\right), & 
 \Delta Q =& \frac{\mu(\bm{\omega}^2 + \bm{a}^2)}{4\pi^2},\nonumber\\
 P_\beta =& P_{\rm F} + \Delta P, &
 P_{\rm F} =& 
 \frac{7\pi^2 T^4}{180} + 
 \frac{T^2 \mu^2}{6} + \frac{\mu^4}{12\pi^2}, &
 \Delta P =& \frac{3\bm{\omega}^2 + \bm{a}^2}{72}\left(T^2 + \frac{3\mu^2}{\pi^2}\right),\nonumber\\
 \sigma^\tau_V =& \frac{\mu}{6\pi^2}, &
 \sigma^\tau_\varepsilon =&
 -\frac{1}{18}
 \left(T^2 + \frac{3\mu^2}{\pi^2}\right), &
 \Pi_\beta =& 0.
 \label{eq:beta_frame}
\end{align}
The above results agree with 
\cite{Buzzegoli:2018,Ambrus:2019ayb,Ambrus:2019khr,Ambrus:2017}.
The first terms in $Q_\beta$ and $P_\beta$ coincide 
with the RKT results in (\ref{eq:RKT_M0}) 
with $g_{\rm {F}} = 2$.
On the rotation axis, where $\rho = 0$, equation \eqref{eq:beta_frame} shows that 
the conductivities $\sigma^\tau_V$ and $\sigma^\tau_\varepsilon$ remain finite, while the circular vector $\tau^{\hat{\alpha}}$ vanishes.
This conclusion holds also in the massive case. 
This can be seen by noting that, according to equations (\ref{eq:CC_aux},  \ref{eq:SET_frame}),
both $\braket{:{\widehat{J}}^{\hat{\varphi}}:}_{T_{0}}$ and $\braket{:\widehat{T}_{\hat{t}\hat{\varphi}}:}_{T_{0}}$ vanish when $\rho = 0$. Furthermore, $E_\beta = \braket{:T_{\hat{t}\hat{t}}:}_{T_{0}}$ (since $\rho\Omega = 0$) and it can be shown that $\braket{:T_{\hat{\rho}\hat{\rho}}:}_{T_{0}} = \braket{:T_{\hat{\varphi}\hat{\varphi}}:}_{T_{0}} = \braket{:T_{\hat{z}\hat{z}}:}_{T_{0}}$ and thus, the SET takes the perfect fluid form at $\rho= 0$. 

\subsection{Quantum corrections to the SET}

We now examine the effect of quantum corrections on the SET, comparing first the exact RKT results (\ref{eq:RKT_M0}) and QFT results \eqref{eq:beta_frame} in the massless case. 
There are three features of note.

First, quantum corrections mean that the SET no longer has the perfect fluid form, due to the presence of nonequilibrium terms, except on the axis of rotation, where 
the circular vector $\tau^{\hat{\alpha}}$ \eqref{eq:kinematic} vanishes.
Second, the quantum corrections to the equilibrium quantities $Q_{\beta }$, $E_{\beta }$ and $P_{\beta }$ are proportional to $\Omega ^{2}$.
Third, the quantum corrections in \eqref{eq:beta_frame} diverge more quickly than the RKT quantities as $\Gamma \rightarrow \infty $ and the SLS is approached.
Therefore there is a neighbourhood of the SLS where quantum corrections become dominant. 

In order to assess the relative contribution made by
quantum corrections with respect to the RKT results,
we first focus on the energy density for massless 
particles and consider two quantities (we restore the reduced Planck's constant $\hbar $ and Boltzmann constant $k_{\rm {B}})$:
\begin{align}
\frac{E_{\beta}}{E_F} - 1 =& \frac{15}{14} 
\left(\frac{\hbar \Omega}{\pi k_{\rm {B}} T_0}\right)^2 \left(\frac{4}{3}
\Gamma^2 - \frac{1}{3}\right) 
\frac{1 + 3 (\mu_0/ \pi k_{\rm {B}} T_0)^2}
{1 + \frac{30}{7} (\mu_0 / \pi k_{\rm {B}} T_0)^2 +
\frac{15}{7} (\mu_0 / \pi k_{\rm {B}} T_0)^4},
\nonumber\\
1 - \frac{E_F}{E_{\beta}} =& 
\left[1 + \frac{14}{5(4\Gamma^2 - 1)} 
\left(\frac{\pi k_{\rm {B}} T_0}{\hbar \Omega}\right)^2 
\frac{1 + \frac{30}{7} (\mu_0 / \pi k_{\rm {B}} T_0)^2 + 
\frac{15}{7} (\mu_0 / \pi k_{\rm {B}} T_0)^4}
{1 + 3 (\mu_0 / \pi k_{\rm {B}} T_0)^2}\right]^{-1}.
\label{eq:Erel}
\end{align}

\begin{figure}[h]
\centering
\begin{tabular}{cc}
 \includegraphics[width=0.45\linewidth]{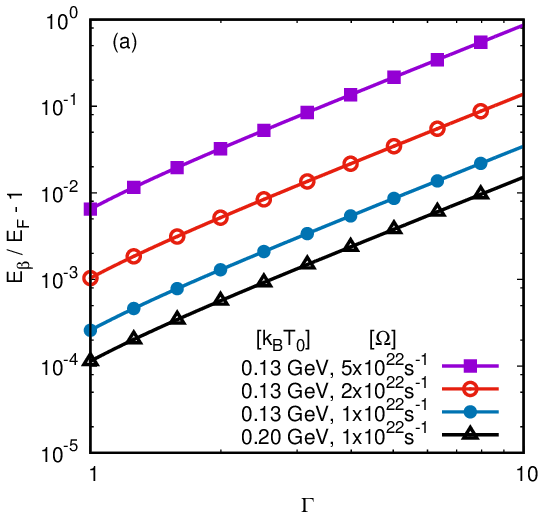} &
 \includegraphics[width=0.47\linewidth]{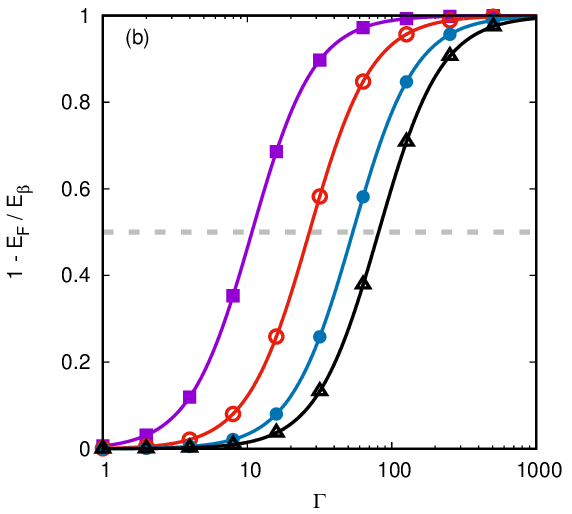} 
\end{tabular}
\caption{Relative differences (a) $E_\beta / E_F - 1$ 
and  (b) $1 - E_F /E_\beta$ between the $\beta$-frame 
energy density $E_\beta$ \eqref{eq:beta_frame} and the RKT result $E_F$ \eqref{eq:RKT_M0} for massless fermions.
The curves correspond to $k_{\rm {B}} T_0 = 0.13 \ {\rm GeV}$
(filled purple squares, empty red circles and filled 
blue circles) and $0.2\  {\rm GeV}$ (empty black 
triangles). The angular velocity is set to 
$\Omega = 5 \times 10^{22}\,{\rm s}^{-1}$ (filled 
purple squares), $2 \times 10^{22}\,{\rm s}^{-1}$
(empty red circles) and $10^{22}\,{\rm s}^{-1}$
(filled blue circles and empty black triangles).
The chemical potential on the rotation axis 
is  $\mu_0 = 0.1\ {\rm GeV}$.
}
\label{fig:QFT_M0}
\end{figure}

Figure~\ref{fig:QFT_M0}(a) shows the relative 
departure of the QFT energy density $E_\beta$ \eqref{eq:beta_frame}
measured in the thermometer frame, 
compared to the RKT energy density $E_F = 3P_F$ \eqref{eq:RKT_M0}. 
We use values of the chemical potential and angular speed relevant for heavy ion collisions, as in section~\ref{sec:RKT:macro}.
For $k_{\rm {B}} T_0 = 0.2\ {\rm GeV}$ and $\Omega = 10^{22}\,{\rm s}^{-1}$, the 
relative difference is about $10^{-4}$ on the rotation axis. 
From \eqref{eq:Erel}, this value can be 
increased by either increasing the angular velocity $\Omega $
or decreasing the temperature $T_{0}$. 
We thus also consider 
a lower temperature relevant to the QGP,
$k_{\rm {B}}T_{0}\simeq 0.13\ {\rm GeV}$.
This enlarges the relative difference by a factor of $\sim 2.4$.
At larger values of the angular speed,
quantum corrections are close 
to $1\%$ on the rotation axis.
Away from the rotation axis,  
the relative difference $E_\beta / E_F - 1$
increases roughly as 
$\Gamma^2 $ \eqref{eq:Erel}. 
This  is confirmed for all regimes considered in 
figure~\ref{fig:QFT_M0}(a).

The  relative difference $1 - E_F / E_\beta$ is presented in figure~\ref{fig:QFT_M0}(b).
On the rotation axis, this ratio is negligible. 
As 
$\Gamma\rightarrow \infty$, 
equation \eqref{eq:Erel} shows that the second term 
in the square bracket goes to $0$ and thus 
$\lim_{\Gamma \rightarrow \infty} 
1 - E_F / E_\beta \rightarrow 1$. 
Close 
to the SLS, quantum corrections therefore become the 
dominant contribution to the energy density $E_\beta$.
The gray, dashed line in figure~\ref{fig:QFT_M0}(b) indicates
where the quantum corrections become equal to
the classical contribution, $E_\beta =2E_F$.
This happens 
closer to the SLS when the temperature is increased
or when the angular velocity is decreased.

\begin{figure}
\centering
\begin{tabular}{cc}
 \includegraphics[width=0.46\linewidth]{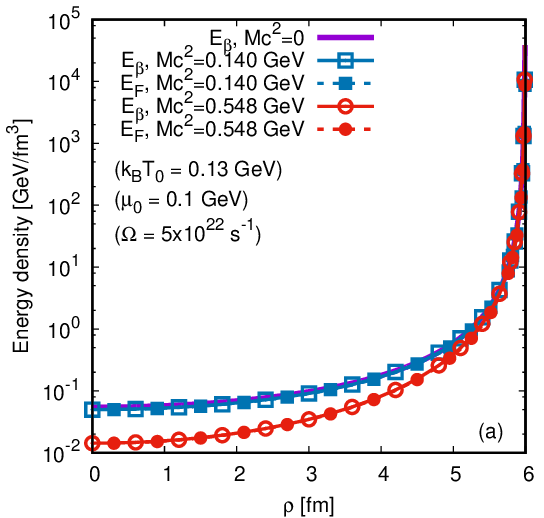} &
 \includegraphics[width=0.45\linewidth]{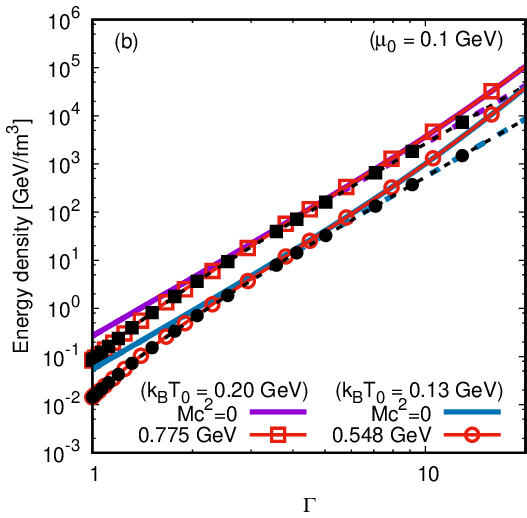} 
\end{tabular}
\caption{Dependence on (a) the distance $\rho$, 
measured in ${\rm fm}$ from the rotation axis,
and (b) on the Lorentz factor $\Gamma$ \eqref{eq:RKT_sol},
of the energy densities $E_\beta$ and $E_F$ obtained 
in QFT (empty symbols and continuous lines)
and RKT (filled symbols and dashed lines) 
at $\mu_0 = 0.1\ {\rm GeV}$
and $\Omega = 5 \times 10^{22}\,{\rm s}^{-1}$.
In (a), the temperature on the rotation axis is fixed 
at $k_{\rm {B}} T_0 = 0.13\ {\rm GeV}$ and the mass 
$Mc^2$ is set to $0$ (continuous purple line,
only $E_\beta$ is shown), $0.140\ {\rm GeV}$ 
(blue squares) and $0.548\ {\rm GeV}$ (red circles). 
In (b), $k_{\rm {B}} T_0 = 0.20\ {\rm GeV}$ (upper lines) 
and $0.13\ {\rm GeV}$ (lower lines). The analytic results
for the massless limit are shown using continuous
(QFT) and dashed (RKT) lines without symbols (purple
is used for $k_{\rm {B}} T_0 = 0.2\ {\rm GeV}$ and 
blue corresponds to $k_{\rm {B}} T_0 = 0.13\ {\rm GeV}$).
}
\label{fig:QFT_Ebeta}
\end{figure}

We next consider the effect of the mass on the energy
density $E_\beta$. 
Figure~\ref{fig:QFT_Ebeta}(a) shows a comparison 
between the energy densities $E_\beta$ and $E_F$, 
as functions
of the distance $\rho$ from the rotation axis.
When 
$\Omega = 5 \times 10^{22}\,{\rm s}^{-1}$, the SLS is located at $\rho = c / \Omega = 6\ {\rm fm}$.
The energy density for particles of mass $ 0.14\ {\rm GeV}$ follows the result for the massless limit very closely, 
while the case with $Mc^2 = 0.548\ {\rm GeV}$ 
can be distinguished from the massless limit only up to 
$\rho \lesssim 5.5\ {\rm fm}$.
Figure~\ref{fig:QFT_Ebeta}(b) shows the dependence of the energy densities 
$E_\beta$ and $E_F$ on the Lorentz factor 
$\Gamma $
\eqref{eq:RKT_sol}.
The RKT and QFT energy densities
can be distinguished when 
$\Gamma \gtrsim 10$, where the higher order 
divergence induced by the quantum corrections
becomes important. 
At large values of $\Gamma$, both the QFT and RKT energy densities follow their respective massless asymptotics, indicating that 
also in the QFT case, the corrections due to 
the mass terms contribute at a subleading order 
close to the SLS, compared with the corresponding 
massless limit.

\begin{figure}
\centering
\begin{tabular}{ccc}
 \includegraphics[width=0.32\linewidth]{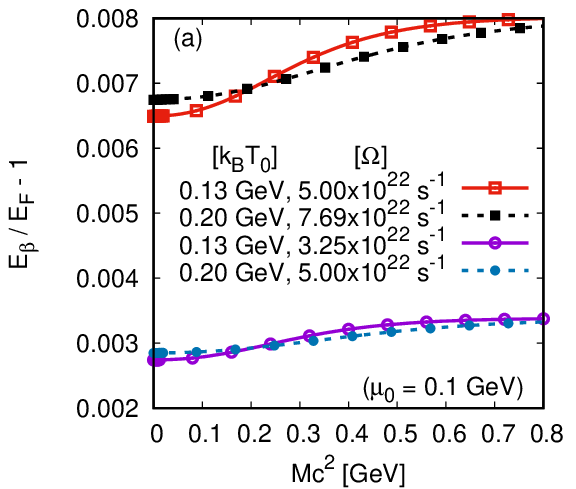} &
 \includegraphics[width=0.32\linewidth]{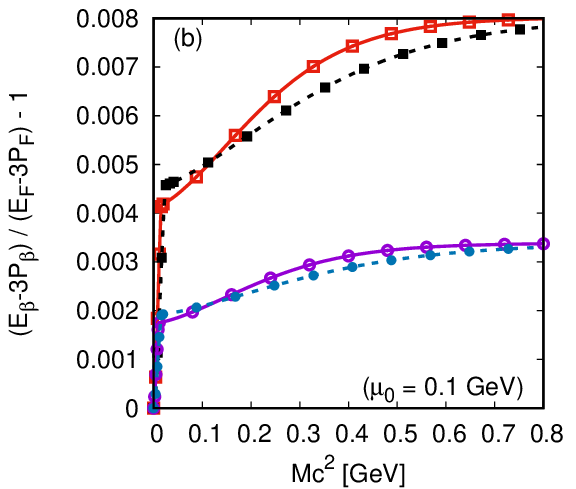} &
 \includegraphics[width=0.32\linewidth]{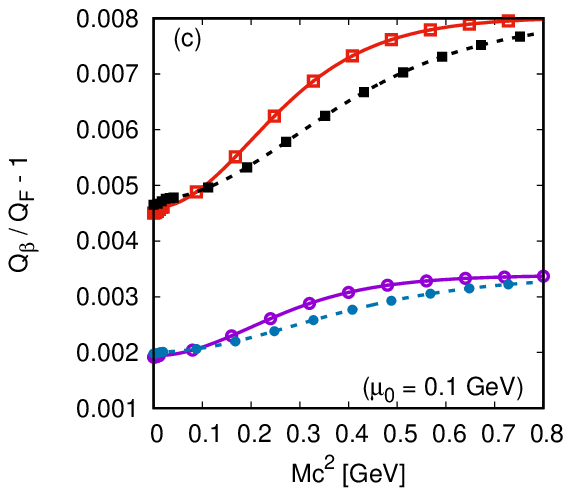} 
 \end{tabular}
\caption{Relative differences
(a) $E_\beta / E_F - 1$, 
(b) $(E_\beta-3P_\beta)/(E_F-3P_F) - 1$,  
(c) $Q_\beta / Q_F -1$, on the 
rotation axis ($\rho = 0$), as functions of the particle mass.
The chemical potential on the 
rotation axis is $\mu_0 = 0.1\ {\rm GeV}$, and 
the temperature on the rotation axis is set to 
$k_{\rm {B}} T_0 = 0.13\ {\rm GeV}$ (empty symbols and 
continuous lines) and $0.2\ {\rm GeV}$ (filled symbols
and dashed lines). 
We consider angular speeds $\Omega $ 
equal to $5 \times 10^{22}\,{\rm s}^{-1}$ 
(red empty squares with continuous lines and 
filled blue circles with dashed lines), 
$7.70 \times 10^{22}\, {\rm s}^{-1}$ (black 
filled squares with dashed lines) and 
$3.25 \times 10^{22}\,{\rm s}^{-1}$ (purple 
empty circles with continuous lines).
}
\label{fig:QFT_axis}
\end{figure}

Finally, we discuss the properties of quantum corrections
on the rotation axis. Since the nonequilibrium 
terms vanish on the rotation axis, only the 
equilibrium quantities, $E_\beta$, $P_\beta$ and 
$Q_\beta$ need to be considered (we assume that $\varpi =0$ here). Instead of discussing
$P_\beta$, we focus on the trace of the SET. 
Figure~\ref{fig:QFT_axis} shows the properties of the
quantum corrections  (a) $E_\beta / E_F - 1$, 
(b) $(E_\beta-3P_\beta)/(E_F-3P_F) - 1$ and 
(c) $Q_\beta / Q_F -1$, computed as relative differences
between the QFT and RKT results. 

Focussing on the small mass regime, it can 
be seen that the relative quantum corrections of the SET trace
exhibit a rapid variation with respect to 
$M$. This variation can be 
attributed to the presence of the sign function 
in the SET components \eqref{eq:SET_frame}, which can take negative
values only when $Mc^2 < \hbar \Omega / 2$. 
In particular, the quantity 
$(E_\beta - 3P_\beta)/M^2 c^4$ exhibits no quantum 
corrections with respect to the corresponding RKT 
quantity when $M = 0$. A rapid increase can be 
seen at small masses bringing the relative quantum 
corrections to the SET trace from zero to the values observed 
for the other quantities (energy and charge density).
At intermediate masses, a slow increase in the 
relative quantum corrections of all quantities can be seen. 
In the large mass limit, the relative quantum 
corrections seem to  reach a plateau value. 

\section{Rigidly-rotating quantum systems in curved space-time}
\label{sec:further}

Thus far, we have focussed our attention on a quantum field in a rigidly-rotating state on unbounded Minkowski space-time.
We have seen that thermal states for such a set-up cannot be defined if the quantum field is a scalar field \cite{Vilenkin:1980,Duffy:2002ss}.
However, it is possible to define rigidly-rotating thermal states for a quantum scalar field constrained within a cylindrical reflecting boundary enclosing the axis of rotation, providing the boundary lies completely within the SLS \cite{Vilenkin:1980,Duffy:2002ss}.
In this latter situation the rotating vacuum is identical to the nonrotating vacuum state and  t.e.v.s are well-behaved.
In \cite{Duffy:2002ss} it is shown that the t.e.v.s in a corotating frame are very well approximated by the RKT quantities derived in section~\ref{sec:RKT}, except for a region close to the boundary, where the Casimir effect becomes important.

In this chapter we have shown that the situation on unbounded Minkowski space-time is very different for a fermion field compared to a scalar field \cite{Ambrus:2014uqa}, in particular we can define a rotating fermion quantum vacuum state and rigidly-rotating thermal fermion states.
T.e.v.s in these states are regular up to the SLS, where they diverge.
A natural question is whether it is possible to consider a set-up similar to that for the scalar field, namely by including a reflecting boundary.
For fermions, defining reflecting boundary conditions is more involved than it is for scalars (where one can simply impose, for example, Dirichlet boundary conditions).  
Using either nonlocal spectral boundary conditions \cite{Hortacsu:1980kv}
or the local MIT-bag boundary condition \cite{Chodos:1974je} on a cylindrical boundary inside the SLS, the rotating fermion vacuum is identical to the nonrotating fermion vacuum \cite{Ambrus:2015lfr}.
Furthermore, rigidly-rotating thermal states have well-defined t.e.v.s, which are computed in \cite{Ambrus:2015lfr} for the case of zero chemical potential. 
At sufficiently high temperatures, the t.e.v.s for the bounded scenario are very well approximated by the unbounded t.e.v.s we have discussed in sections~\ref{sec:tevs} and~\ref{sec:SET}, except for a region close to the boundary.
In \cite{Chernodub:2017mvp} it is shown that, as well as the ``bulk'' mode considered in \cite{Ambrus:2015lfr}, the fermion field also has ``edge states'' localized near the boundary, which must also be taken into account. 
The effect of interactions for rigidly-rotating fermions inside a cylindrical boundary is studied in \cite{Chernodub:2016kxh,Chernodub:2017ref}.

In Minkowski space-time a rigidly-rotating quantum system is therefore unphysical unless an arbitrary boundary is introduced in such a way that there is no SLS.
A natural question is whether rigidly-rotating quantum states exist in curved space-time.
One advantage of working on Minkowski space-time is that, as well as having no curvature, the space-time has maximal symmetry, which simplifies many aspects of the analysis.
To explore the effect of space-time curvature on rigidly-rotating quantum states, one may consider anti-de Sitter space-time (adS) \cite{Hawking:1973,Moschella2005}. 
This space-time has maximal symmetry but constant negative curvature.
Furthermore, the boundary of the space-time is time-like, as is a cylindrical boundary in Minkowski space-time.
In particular, appropriate conditions have to be applied to the field on the space-time boundary \cite{Avis:1977yn}.

The properties of nonrotating thermal states on adS have been studied in the framework of RKT and QFT, for both scalars \cite{Ambrus:2018olh} and fermions  \cite{Ambrus:2018olh,Ambrus:2017cow}, in the absence of a chemical potential.
The curvature of adS space-time affects these states in a number of ways.
First, the normal-ordering procedure applied in section~\ref{sec:tevs0} is not valid in a general curved space-time due to the fact that v.e.v.s for the nonrotating vacuum are nonzero, for both scalars \cite{Kent:2014nya} and fermions \cite{Ambrus:2015mfa}. 
Unlike our Minkowski space-time results in section~\ref{sec:tevs0}, the t.e.v.s for stationary states of both scalars and fermions receive quantum corrections in adS \cite{Ambrus:2018olh,Ambrus:2017cow,Ambrus:2017vlf}.

What about rigidly-rotating quantum states in adS?  
Due to its time-like boundary, there is no SLS in adS if $\Omega {\mathcal {R}}<1$, where $\Omega $ is the angular speed and ${\mathcal {R}}$ is the radius of curvature of the space-time.
In other words, if the radius of curvature is small and the angular speed not too large, there is no SLS.  
Rigidly-rotating quantum states on adS have been studied in much less detail than their Minkowski counterparts.
For a quantum scalar field, it is known that the only possible choice of global vacuum state is the nonrotating vacuum \cite{Kent:2014wda}, as in Minkowski space-time.
One might conjecture that rigidly-rotating thermal states for scalars can be defined only if there is no SLS, but this question has yet to be addressed.
For a quantum fermion field, 
the rotating and nonrotating vacua are identical if there is no SLS, while if an SLS is present, a distinct rotating vacuum state can be defined
\cite{Ambrus:2014fka}.
The preliminary analysis in \cite{Ambrus:2014fka} shows that rigidly-rotating thermal states have at least some features similar to those seen in sections~\ref{sec:tevs} and \ref{sec:SET} in Minkowski space-time, in particular the t.e.v.s diverge on the SLS (if there is one).

These results demonstrate that space-time curvature does have an effect on rigidly-rotating quantum states.
Asymptotically-adS space-times in particular may be relevant for studying the QGP via gauge-gravity duality (see, for example,  \cite{CasalderreySolana:2011us,DeWolfe:2013cua,Aharony:1999ti,Ammon:2015wua}  for reviews).
In this approach, string theory on an asymptotically adS space-time is dual to a conformal quantum field theory (CFT) on the boundary of adS (which itself is conformal to Minkowski space-time).
The idea is that calculations on one side of the duality may shed light on phenomena on the other side.
For example, thermal states in the boundary CFT would correspond to asymptotically adS black holes in the bulk.
This is because black holes emit thermal quantum radiation \cite{Hawking:1974sw}, the temperature of the radiation being known as the Hawking temperature.
Asymptotically adS rotating black holes \cite{Carter:1968ks} can be in thermal equilibrium with radiation at the Hawking temperature provided either the black hole rotation is not too large, or the adS radius of curvature is sufficiently small \cite{Hawking:1998kw}.
These conditions ensure that there is no SLS for these black holes.
A full QFT computation of the t.e.v.~of the stress-energy tensor for a quantum field on a rotating asymptotically adS black hole is, however, absent from the literature.

Some of the most astrophysically important space-times with rotation are Kerr black holes 
\cite{Kerr:1963ud}.
These black holes are asymptotically flat, that is, far from the black hole the space-time approaches Minkowski space-time, rather than adS space-time as for the black holes discussed in the previous paragraph.
Kerr black holes therefore always have an SLS, a surface on which an observer must travel at the speed-of-light in order to corotate with the black hole's event horizon.
The quantum state describing a black hole in thermal equilibrium with radiation at the Hawking temperature is known as the Hartle-Hawking state \cite{Hartle:1976tp}.
In contrast to the situation for asymptotically adS rotating black holes, such a state cannot be defined for a quantum scalar field on an asymptotically flat Kerr black hole \cite{Kay:1988mu,Ottewill:2000qh}.
Indeed, it can be shown that any quantum state which is isotropic in a frame  rigidly-rotating with the event horizon of the black hole must be divergent at the SLS \cite{Ottewill:2000yr}.
If the black hole is enclosed inside a reflecting mirror sufficiently close to the event horizon of the black hole, then a Hartle-Hawking state can be defined for a quantum scalar field \cite{Duffy:2005mz}.
Interestingly, this state is not exactly rigidly-rotating with the angular speed of the horizon \cite{Duffy:2005mz}. 
For a quantum fermion field, it is possible to define a Hartle-Hawking-like state on the Kerr black hole without the mirror present \cite{Casals:2013}.  While this state is also not exactly rigidly-rotating, it is nonetheless divergent on the SLS \cite{Casals:2013}. 

Rotating black hole space-times are much more complicated that the toy model of rigidly-rotating states on Minkowski space-time that we consider in this chapter.
However, the key physics remains the same in both situations.
Namely, rigidly-rotating states cannot be defined for a quantum scalar field if there is an SLS present.
Rigidly-rotating thermal states can be defined for a quantum fermion field, even when there is an SLS, but such states diverge as the SLS is approached.

\section{Summary}
\label{sec:conc}

In this chapter we have considered the properties of rigidly-rotating systems in QFT.
Our toy models are free massive scalar and fermion fields on unbounded flat space-time.
Such systems cannot be realized in nature due to the presence of the SLS, the surface outside which particles must travel faster than the speed of light in order to be rigidly rotating.
Nonetheless, this approach has revealed some interesting physics which is relevant to more realistic set-ups, such as the QGP as formed in heavy-ion collisions or quantum fields on black hole space-times.

We began the chapter by briefly reviewing the properties of rigidly-rotating thermal states for scalar and fermion particles within the framework of RKT.  
The main feature is that, for both scalars and fermions, macroscopic quantities such as the energy and pressure diverge on the SLS but are regular inside it. 

Next we constructed rigidly-rotating thermal states within the canonical quantization approach to QFT on unbounded Minkowski space-time.
Here there is a significant difference between scalar and fermion fields.
In particular, rigidly-rotating thermal states for scalars cannot be defined.
The quantization of the fermion field is less constrained than that of the scalar field, and as a result we are able to define rigidly-rotating thermal states for fermions.
We computed the t.e.v.s of the FC, CC, AC and SET in these states.
All t.e.v.s diverge on the SLS but are regular inside it.
Relative to the RKT results, the quantum t.e.v.s diverge more rapidly as the SLS is approached. 
Quantum corrections therefore dominate close to the SLS.
We stress that the advantage of the canonical quantization approach considered in this chapter is that 
it allows t.e.v.s to be expressed in integral form, which can then be used to obtain analytic (in the massless case) or numerical (in the massive case) results in a non-perturbative fashion, with arbitrary numerical precision, even in the regime where quantum corrections are dominant.

The toy model considered in this chapter is a good approximation to more physical rigidly-rotating systems enclosed inside a reflecting boundary, except in the vicinity of the boundary. 
The key physics features are also shared with more complicated systems in curved space-time.
We therefore conclude that our method based on canonical quantization can serve as a reliable tool to compute t.e.v.s in rigidly-rotating systems of particles, in particular in set-ups relevant to relativistic heavy-ion collisions,
from the nearly-classical regime to the quantum-dominated regime, with arbitrary numerical precision.

\begin{acknowledgement}
The work of V.~E.~Ambru\cb{s} is supported by a grant from the Romanian 
National Authority for Scientific Research and Innovation, CNCS-UEFISCDI,
project number PN-III-P1-1.1-PD-2016-1423. The work of E.~Winstanley 
is supported by the Lancaster-Manchester-Sheffield Consortium for Fundamental 
Physics under STFC grant ST/P000800/1 and partially supported by the 
H2020-MSCA-RISE-2017 Grant No. FunFiCO-777740.
\end{acknowledgement}

\end{document}